\documentclass[preprint2]{aastex6}
\usepackage{epstopdf}
\usepackage{graphicx}
\usepackage{subfigure}
\usepackage{xcolor}

\usepackage{amsmath}
\usepackage{amssymb}
\usepackage{wasysym}
\usepackage{natbib}

\graphicspath{ {./} }
\DeclareGraphicsExtensions{.pdf,.tikz,.eps}

\renewcommand{\eqref}[1]{eq. \ref{#1}}

\newcommand\gas{\mathrm{gas}}
\newcommand\dust{\mathrm{dust}}

\begin{document}



\title{Self-Sustained Recycling in the Inner Dust Ring of Pre-Transitional Disks}


\author{T. Husmann\altaffilmark{1}, C. Loesche, G. Wurm}

\affil{Fakult{\"a}t f{\"u}r Physik, Universit{\"a}t Duisburg-Essen\\
Lotharstr. 1\\
47048 Duisburg, Germany}

\altaffiltext{1}{tim.jankowski@uni-due.de}

\begin{abstract}

Observations of pre-transitional disks show a narrow inner dust ring and a larger outer one.  They are separated by a cavity with no or only little dust. We propose an efficient recycling mechanism for the inner dust ring which keeps it in a steady-state. No major particle sources are needed for replenishment. Dust particles and pebbles drift outwards by radiation pressure and photophoresis. The pebbles grow during outward drift until they reach a balanced position where residual gravity compensates photophoresis. While still growing larger they reverse their motion and drift inwards. Eventually, their speed is fast enough that they get destroyed in collisions with other pebbles and drift outwards again. We quantify the force balance and drift velocities for the disks LkCa15 and HD135344B. We simulate single particle evolution and show that this scenario is viable. Growth and drift timescales are on the same order and a steady state can be established in the inner dust ring. 

\end{abstract}


\section{Introduction} \label{sec:intro}

Protoplanetary disks with optically thin gaps (pre-transitional disks) have been observed frequently \citep{Calvet2002,Najita2007,Sicilia-Aguilar2008, Bruderer2014,Marel2015}. The basic structure of these disks is as follows.
\begin{itemize}
\item There is an inner disk up to a gap opening distance from the star $r_{\rm gap}$ which is usually on the order of about one AU.  
\item This inner ring contains dust in a significant amount. Radiation from the central star can cross the inner dust disk depending on the dust density but with intensity loss up to several orders of magnitude. 
\item There is a cavity from $r_{\rm gap}$ to $r_{\rm cav}$, where the disk contains (nearly) no dust. 
\item The inner edge of the outer disk $r_{\rm cav}$ is usally several tens of AU from the star.  
\item The inner disk up to $r_{\rm cav}$ is not gas-free. As accretion partly goes on and as recent CO measurements show, the gas content can be very significant.
\item Compared to the gas the solid fraction of the disk is strongly reduced. This explicitly includes small sub-micron dust in the warmer inner parts of the gap.
\end{itemize}
Considering this basic setup, one would expect the inner dust ring to vanish quickly since there is no dust reservoir outside the ring to drift inwards and replenish particle losses that might be assumed due to accretion or particle growth. 

In this paper we introduce a recycling mechanism which takes place in the inner dust ring and keeps a self-sustained dust distribution. It prevents particle accretion by the star and it prevents dust to vanish due to unlimited growth to larger sizes no longer observable. Classic calculations of dust movement (e.g. by \cite{Weidenschilling1977a} which is adopted frequently) do not include additional radial forces. Therefore in our model we included radiation pressure and photophoretic forces. 
Photophoresis has long been known and applied in atmospheric science \citep{Rohatschek1995,Cheremisin2005,Beresnev2003c}. It was only introduced
to protoplanetary disks by \citet{Krauss2005} and \citet{Wurm2006} as potential candidate to drive
dust particles and chondrules and concentrate them in specific locations. Since then the quantiative description of photophoresis has been improved based on numerical simulations and laboratory experiments  \citep{Wurm2010,VonBorstel2012, Loesche2014, Matthews2016, Loesche2016b}. Futher applications of photophoresis to the inner edge of full protoplanetary disks considered the separation of different materials or local concentration \citep{Haack2007, Wurm2013, Cuello2016}.  Also large scale particle transport by photophoresis in an optical thin disk or over the surface of the disk have been calculated, i.e. to explain how high temperature minerals forming close to the sun can be found further out in the disk, for example in asteroids or comets \citep{Wurm2009, Mousis2007, Moudens2011a}. 
Currently, photophoresis is being explored as source of particle motion in a disk with temperature fluctuations (Loesche et al. submitted) which is work ongoing. Photophoresis by thermal radiation might also be important in the context of (hot) giant planet accretion \citep{Teiser2013}. Related to this more accurate treatments of photophoretic forces in radiation fields of optical thick disks have already been worked out \citep{McNally2015}.
The motion of dust in optical thin disks explicitly includes transitional disks which were considered by \cite{Takeuchi2008} and \cite{Herrmann2007}.
\cite{Dominik2011} looked for an explanation why the gap is so clean of dust. They studied the influence of radiation pressure which pushes the dust outwards but come to the conclusion that this is not a viable mechanism for keeping the gap clean of small dust coming from the outer disk. They consider photophoresis shortly but discard it due to the fact that photophoretic motion is small for sub micron particles. We do not consider in detail here why dust from the outer disk is not crossing the inner edge of the outer disk. This might be a particle trap due to the pressure drop or an embedded planet. We consider radiation pressure and photophoresis for the inner dust ring here. For one thing we show later that, in fact, photophoresis can dominate the particle movement even for $10^{-6}$ m sized particles at least in the inner parts of the disk. In addition though the small grains are embedded in a cloud of larger grains and
cannot be regarded as isolated. While effects of photophoresis on small grains might be low photophoretic drift is significant for larger grains which in turn influences the small grain fraction.

As detailed below, mm-sized particles can set off with velocities up to 10 m/s close to the star and cross a wide distance easily on a short timescale. If they encounter smaller grains on their way outwards they can partly collect them and grow bigger. Photophoretic drift decreases in efficiency with increasing size and a stability point will be reached while feeding further on any small dust particles. This prevents further outward drift of large grains and of small grains which are intercepted by the large particles. Therefore the gap can be kept clean at its inner edge and the dust will stay in the inner ring. Once the large particles reach a critical size they switch their drift direction and move back inwards. Close to the star they meet their equals in size with high velocities and get destroyed in collisions. This produces some dust and pebble size particles again, which -- once more -- move outwards efficiently, scavenge small dust, get back and so on.

\begin{figure*}
	\figurenum{1}
	\plotone{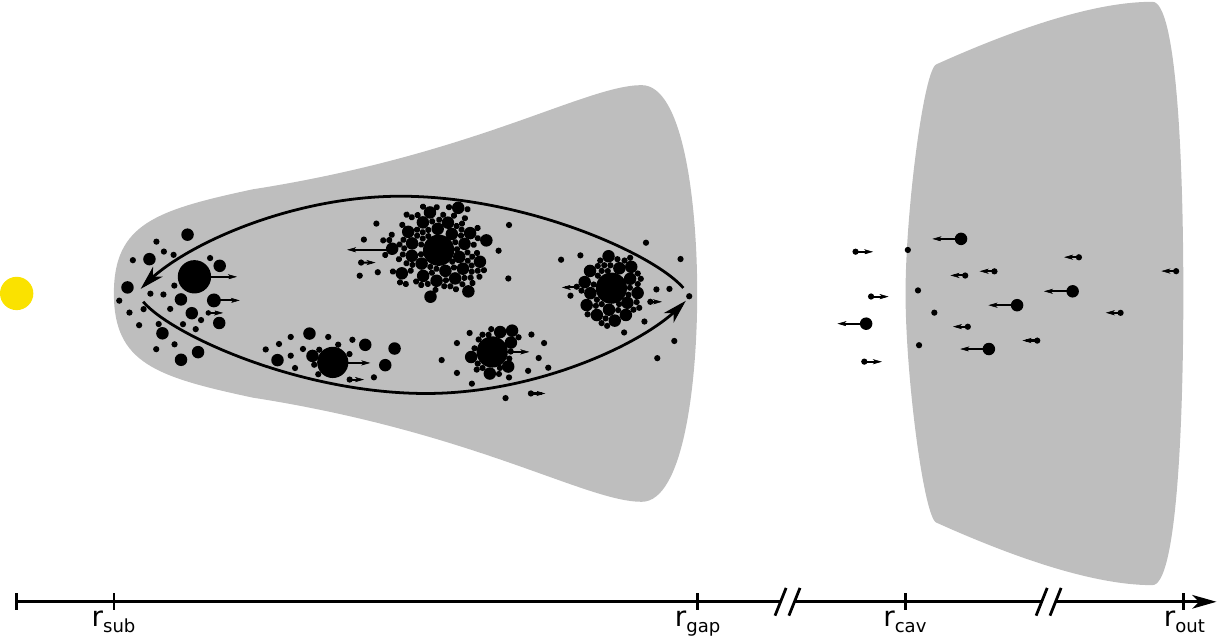}
\caption{Scheme of the disk structure  and sketch of the self-sustained recycling process which keeps the mass inside the inner disk by photophoretic sweep up. Inside the inner disk (between $r_{\rm sub}$ and $r_{\rm cav}$) small particles generally tend to drift outwards. Due to the collisional behaviour, the larger particles sweep up the smaller particles and grow. Once they reach a critical size the drift direction switches and they move inwards which further supports the sweep up process. Eventually, the large particles get destroyed in collisions with particles of (more or less) the same size and the whole process can start again. The disks contains only little dust inside the gap (between $r_{\rm cav}$ and $r_{\rm gap}$) but gas is still present, although less than in the outer disk ($ > r_{\rm gap}$). For small particles the drift is positive for $r \lesssim r_{\rm gap}$ and negative for $r \gtrsim r_{\rm gap}$ while larger particles always drift inwards for large distances.}
\label{fig:diskclearing}
\end{figure*}

This principle of cleaning is actually not unheard of. Cleaning of the Earth's atmosphere works in a similar way. Sub-micron grains (smog) easily accumulate under dry conditions. However, if it rains, eventually, the raindrops
collect the small dust and clean the atmosphere. In that case gravity is the driving force for the large particles instead of photophoresis. The same concept is used technically to clean airflows from small particles as droplets are introduced and capture fine dust.
A sketch of the protoplanetary version of photophoretic sweep up cleaning is shown in fig. \ref{fig:diskclearing}. 

In the following sections we calculate the drift of particles in the disks LkCa15 and HD135344B and perform single particle evolution simulations showing that a self sustained recycling process can be established where large particles drift outwards up to $r_{\rm gap}$, grow beyond a certain mass and fall back inwards where they get destroyed again.

\section{Model Details - Radial Drift Calculations}

\subsection{Disk models}

For our calculations we used two different transitional disks -- LkCa15 (with two different parameter settings) and HD135344B -- with a physical model described by \cite{Andrews2011}. The parameters are adopted from the parameter range given by \cite{Marel2015} and displayed in tab. \ref{table:diskparameters}. As mentioned above, these disks contain a gap between an inner radius $r_\mathrm{gap}$ and an outer radius $r_{\rm cav}$. Furthermore every disk is determined by a set of additional parameters: a critical radius $r_{\rm c}$, the dimensionless scale height at the critical radius $h_c$, the surface density at the critical radius $\Sigma_c$, the flaring angle $\psi$, the gas surface density drop for $r < r_{\rm cav}$ labeled $\delta_\gas$, the dust drop value $\delta_\dust$, the dust scale height $\chi_\dust$, the sublimation radius $r_\mathrm{sub}$, the mass of the star $M$, its effective Temperature $T_{\rm eff}$, and its luminosity $L$. 

The gas surface densitiy $\Sigma_\gas$ and the dust surface density $\Sigma_\dust$ of the disk between $r_\mathrm{sub}$ and $r_\mathrm{gap}$ are described as
\begin{align}
   \Sigma_\gas & = \delta_\gas \cdot \Sigma_\text{c} \left( \frac{r}{r_\text{c}} \right)^{-\gamma} \exp\left(-\frac{r}{r_\text{c}}\right) \\
   \Sigma_\dust & = 
\begin{cases}
\delta_\dust \delta_{\rm DGR} \cdot \Sigma_\gas &  \mathrm{,for } \, r < r_{\rm gap} \\
\delta_\dust \delta_\mathrm{dustcav} \delta_{\rm DGR} \cdot \Sigma_\gas &  \mathrm{,for } \, r_{\rm gap} < r < r_{\rm cav} \\
\delta_{\rm DGR}  \cdot \Sigma_\gas &  \mathrm{,for } \, r > r_{\rm cav} 
\end{cases}
\label{eq:surfacedensity}
\end{align}
For the disks used here $\gamma$ was set to 1 and the general dust to gas mass ratio is $\delta_{\rm DGR} = 0.01$. The dust drop values $\delta_\dust$ and $\delta_{\rm dustcav}$ are dependent on the investigated disk.  $\Sigma_\gas$ is valid even for larger radii up to the edge of the cavity. Note here that we modified the dust and gas drop coefficients slightly according to the parameter range given by \cite{Marel2015}. The scale height is parameterized by
\begin{equation}
   H = h_\text{c}\left(\frac{r}{r_\text{c}}\right)^\psi r \quad .
\label{eq:scaleheight}
\end{equation}

With this in mind and $H = \frac{c_{\rm s}}{\Omega}$ with $c_{\rm s}$ as sound speed  and $\Omega$ as Keplerian frequency, the temperature distribution can be described via
\begin{equation}
   T = T_\mathrm{sub} \left(\frac{r}{r_\mathrm{sub}}\right)^{2\psi-1} \quad .
\label{eq:tgas}
\end{equation}
, with $T_\mathrm{sub}$ is the sublimation temperature of silicate, assumed to be 1500$\,$K \citep[e.g.][]{Baillie2015} and $r_\mathrm{sub}$ is the distance to the central star with $T(r_\mathrm{sub}) = T_\mathrm{sub}$.

From the surface density one can derive the gas and dust density at the midplane by
\begin{align}
\rho_\gas = & \frac{1}{\sqrt{2 \pi}}\frac{\Sigma_\gas}{H} \\
\rho_\dust = & \frac{1}{\sqrt{2 \pi}}\frac{\Sigma_\dust}{\chi_\dust H}
\end{align}
where $\chi_\dust$ is a dimensionless factor reducing the dust scale height $H$ to adapt for dust settling \citep{Dubrulle1995, Andrews2011}. Note here that for our simulations this parameter has the same impact as the dust to gas ratio $DGR$ and the dust drop value $\delta_\dust$. In general, $\chi_\dust$ is dependent on the particle's Stokes number as well as on the strength of the turbulence \citep{Johansen2005, Turner2006}. Assuming a fully turbulent disk, $\chi_\dust$ can get as high as unity. For simplification, we assume $\chi_\dust$ to be constant for every distance $r$ and every particle size $r_{\rm p}$.
The pressure in the midplane is then given by
\begin{equation}
P = \frac{\rho_\gas \, T_\gas \, {\rm R}}{\mu} \quad .
\end{equation}
, with $\mu = 2.3 \times 10^{-3}$ kg mol$^{-1}$ is the molar mass of fully molecular gas of cosmic composition and R is the ideal gas constant.
From the pressure profile the azimuthal gas velocity can be calculated via
\begin{equation}
v_{\phi,\gas}=\sqrt{\frac{G M}{r}+\frac{r \, \frac{\mathrm{d} p}{\mathrm{d} r}}{\rho_\gas}} \quad .
\label{eq:vphigas}
\end{equation}
The radial gas velocity $v_{\rm r,gas}$ is on the order of $10^{-3}$ m/s which is generally much smaller than the radial dust drift $v_{\rm r}$ for the disk LkCa15. Nonetheless, in the disk HD135344B the gas velocity cannot be omitted and will therefore be included in all calculations via \citep{Lyndenbell1974}
\begin{equation}
v_{\rm r,gas} = - \frac{3}{\Sigma_\gas \, \sqrt{r}} \cdot \frac{\partial}{\partial r}\left(\Sigma_\gas \, \nu_\gas \, \sqrt{r} \right)
\label{eq:vrgas}
\end{equation}

Note here that the disk parameters calculated by \cite{Marel2015} are not always fixed values but can be interpreted as orientation values. This is especially -- but not exclusively -- the case for the inner gap radius $r_{\rm gap}$, since \cite{Marel2015} argued that they cannot resolve this radius in detail since changes in this value do not influence the SED data significantly.

\begin{figure}
	\figurenum{2}
	\includegraphics[width=0.45\textwidth]{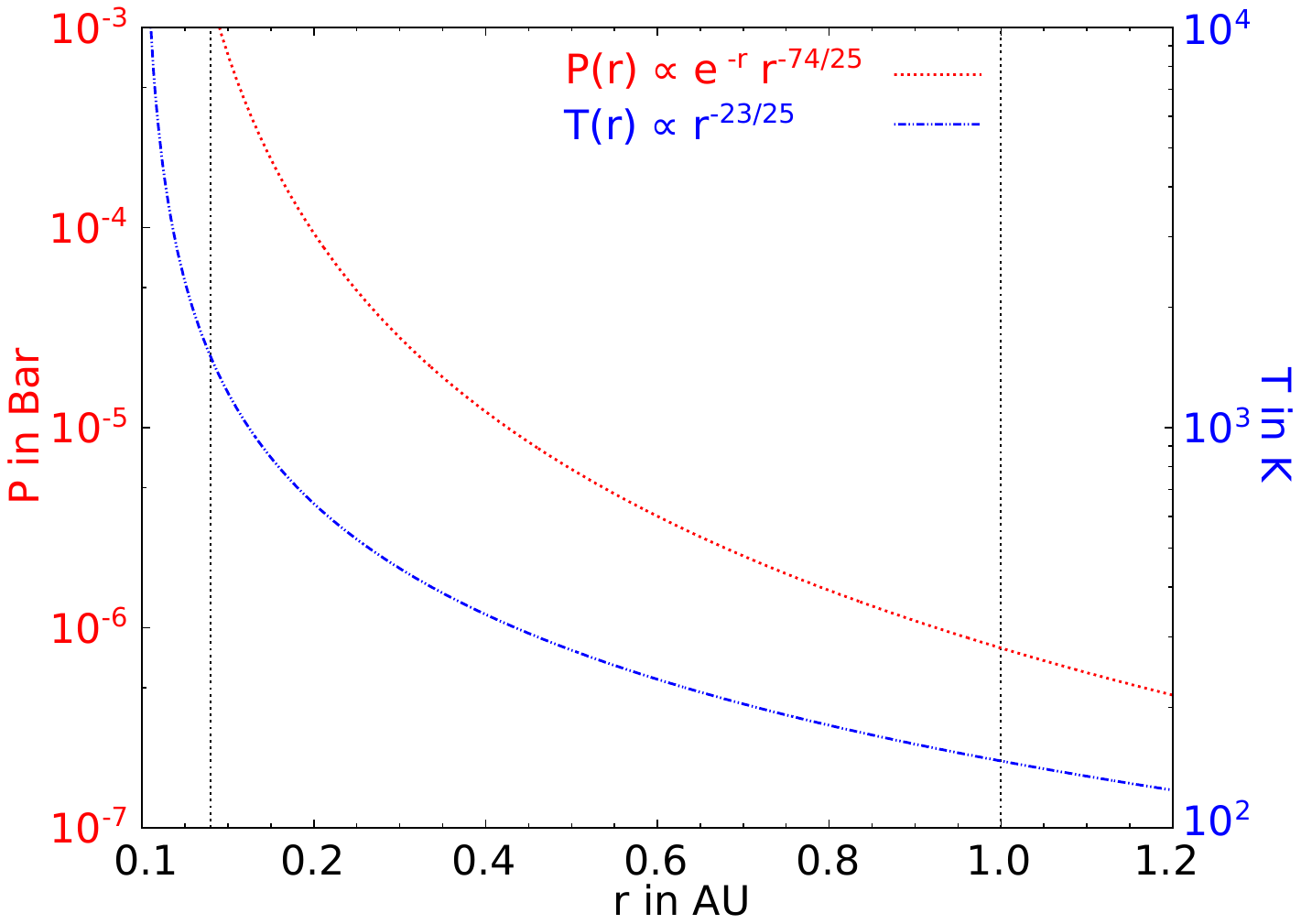}
\caption{Pressure distribution (red, dotted line) and temperature distribution (blue, dashed line) for the disk LkCa15 (Model 1). The dotted black lines denote the sublimentation radius $r_{\rm sub}$ as well as the gap radius $r_{\rm gap}$.}
\label{fig:ptlkca15}
\end{figure}

Since photophoresis as well as radiation pressure is mainly dependent on the intensity of the irradiation, we included an opacity model to calculate the intensity at a given distance via
\begin{equation}
I = I_0 \cdot \exp \left(-\kappa_{\rm med} \int_{r_{\rm sub}}^{r} \rho_\dust (r') \rm{d}r' \right)
\end{equation}
where $I_0$ is the undisturbed initial intensity at a distance r and
\begin{equation}
\kappa_{\rm med} = \frac{\int B_{\rm \lambda} (\lambda, T_{\rm eff}) \kappa_{\rm \lambda} \rm{d}\lambda}{\int B_{\rm \lambda} (\lambda, T_{\rm eff})  \rm{d}\lambda }
\end{equation}
is the mean opacity for the dust and gas for a given star of temperature $T_{\rm eff}$. Since there is not sufficient data to calculate $\kappa_{\rm \lambda}$ for every temperature (and therefore for every distance to the central star), we simplified $\kappa_{\rm \lambda}$ to be equal for every distance to the central star and used the data provided by \cite{Semenov2003} to get an estimation of $\kappa_{\lambda}$. Depending on the silicate mineralogy (iron-rich, iron-poor, normal), the particle type, the temperatures (star and dust/gas), $\kappa_{\rm med}$ can vary between $\approx 1.8$ to $5.5$ m$^2$ kg$^{-1}$. Since most of these parameters are unknown for the investigated disks, we chose $\kappa_{\rm med}$ to be 2.5 (generally hotter temperature of the gas) for the disk HD135344B and 3.0 for LkCa15. As shown later, this is not critical as different disk configurations exist with other values for $\kappa_{\rm med}$ which also form a self-sustained recycling process within the given inner dust disk.
All disk parameters necessary for calculating the particle drift are given in tab. \ref{table:diskparameters}. Note here that two different models for LkCa15 were used.  For the disk LkCa15 (Model 1) the pressure and temperature profiles are plotted in fig. \ref{fig:ptlkca15}. As depicted later (see sec. \ref{subsec:collisionvelocities}) we assume a highly turbulent disk leading to high $\chi_\dust$ values.

\begin{table*}
\caption{Disk parameters adapted from \cite{Marel2015}, note that $\delta_\gas$ and $\delta_\dust$ are slightly modified. M1 denoted the first model used for the disk LkCa15, M2 the second one.}              
\label{table:diskparameters}      
\centering                                      
\begin{tabular}{l l l l l l l l l l l l l l l l}          
\hline\hline                        
Disk 			&	$r_\mathrm{c}$ 	&	$\Sigma_\mathrm{c}$					&	$h_\mathrm{c}$	&	$\psi$	&	$\delta_\gas$	& 	$\delta_\dust$	&	$\kappa_{\rm med}$				&	$r_\mathrm{sub}$	&	$r_\mathrm{gap}$ 	&	$r_\mathrm{cav}$ 	&	$M_\ast$				&	$L_\ast$  				&	$T_{\rm eff}$	& $k_{\rm th}$						&	$\chi_\dust$\\
			&	(AU)			&	$\left( \frac{\rm g}{\rm cm^{2}} \right)$		&	(rad)			&			&				&				&	$\left( \frac{\rm m^2}{\rm kg} \right)$	&	(AU)				&	(AU)				&	(AU)				&	($M_{\astrosun}$)	&	($L_{\astrosun}$)	&	(K)			&	$\left( \frac{\rm W}{\rm m \, K} \right)$ 	&	\\
\hline
HD135344B	&	25		&	300			&	0.15		&	0.05	&	$10^{-2}$			&	$10^{-3}$			&	2.5	&	0.18			&	0.25			&	40		&	1.6	&	7.8 	&	6590	&	$10^{-2}$	& 0.85\\
LkCa15 M1	&	85		&	34			&	0.06		&	0.04	&	$10^{-2}$			&	$2 \times 10^{-5}$	&	3.0	&	0.08			&	1			&	45		&	1.0		&	1.2 	&	4730	&	$10^{-2}$	& 0.5\\
LkCa15 M2	&	85		&	34			&	0.06		&	0.04	&	$5 \times 10^{-3}$	&	$3 \times 10^{-5}$	&	3.0	&	0.08			&	1			&	45		&	1.0		&	1.2 	&	4730	&	$10^{-3}$	& 0.5\\
\hline                                   
\hline
\end{tabular}
\end{table*}

\subsection{Forces on Particles}

Particles moving in a protoplanetary disk are subject to basically four forces (neglecting electrical charges and self-gravity):
\begin{itemize}
\item Gas Drag
\item Radiation Pressure
\item Photophoresis
\item Residual gravity of the star
\end{itemize}

\subsubsection{Gas Drag}

The drag force is described by
\begin{equation}
F_{\rm drag} = \frac{m}{\tau_{\rm f}} \,  \left( v_{\rm p} - v_{\rm gas} \right)
\label{eq:dragforce}
\end{equation}
where $m$ is the particles mass, $v_{\rm p} - v_{\rm gas}$ is the relative velocity between the particle and the surrounding gas and $\tau_{\rm f}$ is the gas-grain friction time. The latter defines how fast a particle couples to the gas. As the particle sizes and pressures in the disks do not allow the calculation of drag force to be either Stokes or Epstein dominated, we used the Cunningham corrected drag formula \citep{Cunningham1910} which interpolates between the free molecular flow regime and the continuum regime. This leads to
\begin{align}
\tau_\text{f} & = \frac{2 \, \rho_{\rm{p}}\,  r_{\rm p}^{2}}{9 \, \eta} C_{\rm cun} \\
C_{\rm cun} & = 1 + \frac{\lambda_{\rm mfp}}{r_{\rm p}} \left( 0.506 \exp \left(-2.0 \,\frac{r_{\rm p}}{\lambda_{\rm mfp}}\right)+1.141 \right)
\end{align}
where $\eta$ is the dynamic viscosity of the gas, $\rho_{\rm p}= \xi \cdot \rho_{\rm s}$ is the particles density ($\xi$ is the filling factor and $\rho_{\rm s}$ is the solid density), and $\lambda_{\rm mfp}$ is the mean free path of the gas molecules. The numerical values used are adopted from \cite{Rader1990} for H$_2$. 

\subsubsection{Photophoresis and Radiation Pressure}

If a particle in a gaseous environment is illuminated, the two light induced forces acting on it are
\begin{itemize}
\item a force directly caused by momentum transport of impinging photons, called radiation pressure ($F_{\rm rp}$) and
\item a force caused by momentum transport of interacting gas molecules, called photophoresis ($F_{\rm phot}$)
\end{itemize}

\textit{Radiation pressure - } radiation pressure can be calculated -- assuming perfect absorption -- by
\begin{equation}
F_\mathrm{rp}=\frac{I}{c}\pi r_{\rm p}^{2},
\end{equation}
where $c$ denotes the speed of light and $r_{\rm p}$ is the particle radius.

\textit{Photophoresis -} the calculation of photophoretic forces $F_{\rm phot}$ follows \cite{Loesche2016b}. The free molecular flow solution (fm)  is given
by \cite{Loesche2016a} which is extended to the continuum (co) in \cite{Loesche2016b}.
\begin{align}
   F^\text{fm}_\text{phot} =  & \frac{\pi}{3} \, \alpha \, \alpha_\text{m} \, \frac{p}{\sqrt{T_\infty^2 + T_\infty \alpha (A_0^{\rm fm} - T_\infty)}} \, r_{\rm p}^2 \nonumber \\ 
						& \cdot \frac{I\,J_1}{\frac{k_{\rm th}}{r_{\rm p}}+h_{\rm fm}+4\sigma_\text{SB}\varepsilon\,\tilde{T}^3} \\
   F^\text{co}_\text{phot} =  & 4\pi\,\kappa_\text{s}\,\frac{\eta^2_\text{dyn}}{\rho_{\rm gas} \, A_0^{\rm co}}\, \frac{I\,J_1}{\frac{k_{\rm th}}{r_{\rm p}} + 2\frac{k_\gas}{r_{\rm p}} + 4\sigma_\text{SB}\varepsilon\,T_{\rm bb}^3}
\label{eq:fphotFMCO}
\end{align}
Here, $\alpha = 0.4$ denotes the thermal (energy) accomodation coefficient, $\alpha_{\rm m} = 1.0$ denotes the momentum accommodation coefficient, $T_\infty$ denotes the gas temperature given by eq. \ref{eq:tgas}, $A_0^{\rm fm}$ and $A_0^{\rm co}$ denote the fm and co solution for the evolution coefficient of the particle temperature, $I$ is the radiation intensity, $J_1 = 1/2$ is a symmetry factor, $k_{\rm th}$ is the thermal conductivity of the particle, $k_{\rm gas}$ the thermal conductivity for the gas, $h_{\rm fm}$ is the heat transfer coefficient in the fm regime, $\kappa_{\rm s} = 1.14$ is the thermal creep coefficient, $\varepsilon = 1$ is the emissivity, $\sigma_{\rm SB}$ is the Stefan-Boltzmann constant, $\tilde{T}$ is the mean particle temperature, $T_{\rm bb} = 3$ K is the background radiation, and $\eta_{\rm dyn}$ is the dynamical gas viscosity.
The co and fm solutions are interpolated following \citep{Rohatschek1995}, leading to a photophoretic force dependent on an optimum pressure $p_{\rm opt}$ via
\begin{equation}
F_{\rm phot} = \frac{2 \, F_{\rm phot}^{\rm max}}{p/p_{\rm opt} + p_{\rm opt}/p} \quad .
\label{eq:fphot}
\end{equation}
The maximum photophoretic force $F_{\rm phot}^{\rm max}$ can be calculated using the approximation made by \cite{Hettner1928}
\begin{equation}
\frac{1}{F_{\rm phot}^{\rm max}} = \frac{1}{F^\text{fm}_\text{phot}} + \frac{1}{F^\text{co}_\text{phot}}
\label{eq:fphotmax}
\end{equation}
We do not depict the calculations in detail here and refer to \cite{Loesche2016a} and \cite{Loesche2016b} for the exact calculation method and the parameters. Nonetheless, the solution shows that $F_{\rm phot}^{\rm max}$ as well as $p_{\rm opt}$ (the optimum pressure) are dependent on the particle radius, the distance to the central star, and the thermal conductivity $k_{\rm th}$. As shown later (fig. \ref{fig:forcecompareLK}), photophoresis can dominate the force balance and an accurate approximation is therefore crucial for drift calculations.

\subsection{Particle Motion}

The solids are defined by their size, density $\rho_{\rm p} = \xi \cdot \rho_{\rm s}$ ($\xi$ is the filling factor and $\rho_{\rm s}$ is the solid density), and a thermal conductivity $k_{\rm th}$.
Particles move radially for the following reasons.
The sub-Keplerian rotation of the gas (eq. \ref{eq:vphigas}) implies an inward drift of the solids
as they lack the pressure support the gas feels. Photophoresis and radiation pressure move 
particles outwards. The actual drift direction and velocity is determined by particle size (and properties) and radial distance to the star.

The absolute motion is calculated for the radial and tangential direction. For transport only the radial direction matters. However, both are important as they influence the collision velocity between 
solid particles.  
It is \citep[see e.g.][]{Takeuchi2002}
\begin{align}
\frac{\rm d}{{\rm d}t} (v_{\rm r}) & =\frac{v_\phi^2}{r}-\frac{G M}{r^2}-\frac{v_{\rm r}-v_{\rm r,\gas}}{\tau_\text{f}}+\frac{F_{\mathrm{rp}}+F_{\mathrm{ph}}}{m} \\
\frac{\rm d}{{\rm d}t} (r \, v_\phi)  & =-\frac{r}{\tau_\text{f}}(v_\phi-v_{\phi,\gas}) \quad .
\label{eq:radazdrift}
\end{align}

For time dependent drift calculations ($v(t), r(t)$), these differential equations were solved numerically for the different disks and particle sizes. The calculations of the distance dependent drift velocities ($v(r)$) can be simplified by assuming the gas grain friction time to be small and solving the steady state solution $\frac{\partial v}{\partial t} = 0$ using the cubic polynomial theorem. 
For the material parameters a solid density of $\rho_{\rm s} = 3000$ kg/m$^3$ and a filling factor of $\xi = 1/3$ were assumed. The accommodation coefficient was taken to be 0.4, the momentum accommodation coefficient and the emissivity to be unity.
We assume the particles to radiate against the microwave background of 3\,K. This neglects the thermal radiation of the surrounding dust and is an approximation but does not change the general picture.

\section{Results for Radial Drift Calculations}

\subsection{Force Balance}

Particles do not move if the total radial force $F_{\rm total}$ vanishes. Without radial gas drag it can be calculated as
\begin{equation}
F_{\rm total} = F_{\rm phot} + F_{\rm rp} + F_{\rm res}.
\label{eq:ftotal}
\end{equation}
This is otherwise equal to eq. \ref{eq:radazdrift} and for particles with coupling times smaller than the orbital timescale the centrifugal and gravity part can also be simplified as
a residual gravity given by \cite{Weidenschilling1977a} as
\begin{equation}
F_{\rm res} = \frac{m}{ \rho_{\gas}} \, \frac{{\rm d}P}{{\rm d}r} \qquad .
\label{eq:fres}
\end{equation}
In general particles will drift radially towards these stability points where the forces equal
\citep{Krauss2005}. However, at all times the particles are subject to collisional evolution.

\subsection{Radial Drift}

\begin{figure}
	\figurenum{3}
	\includegraphics[width=0.45\textwidth]{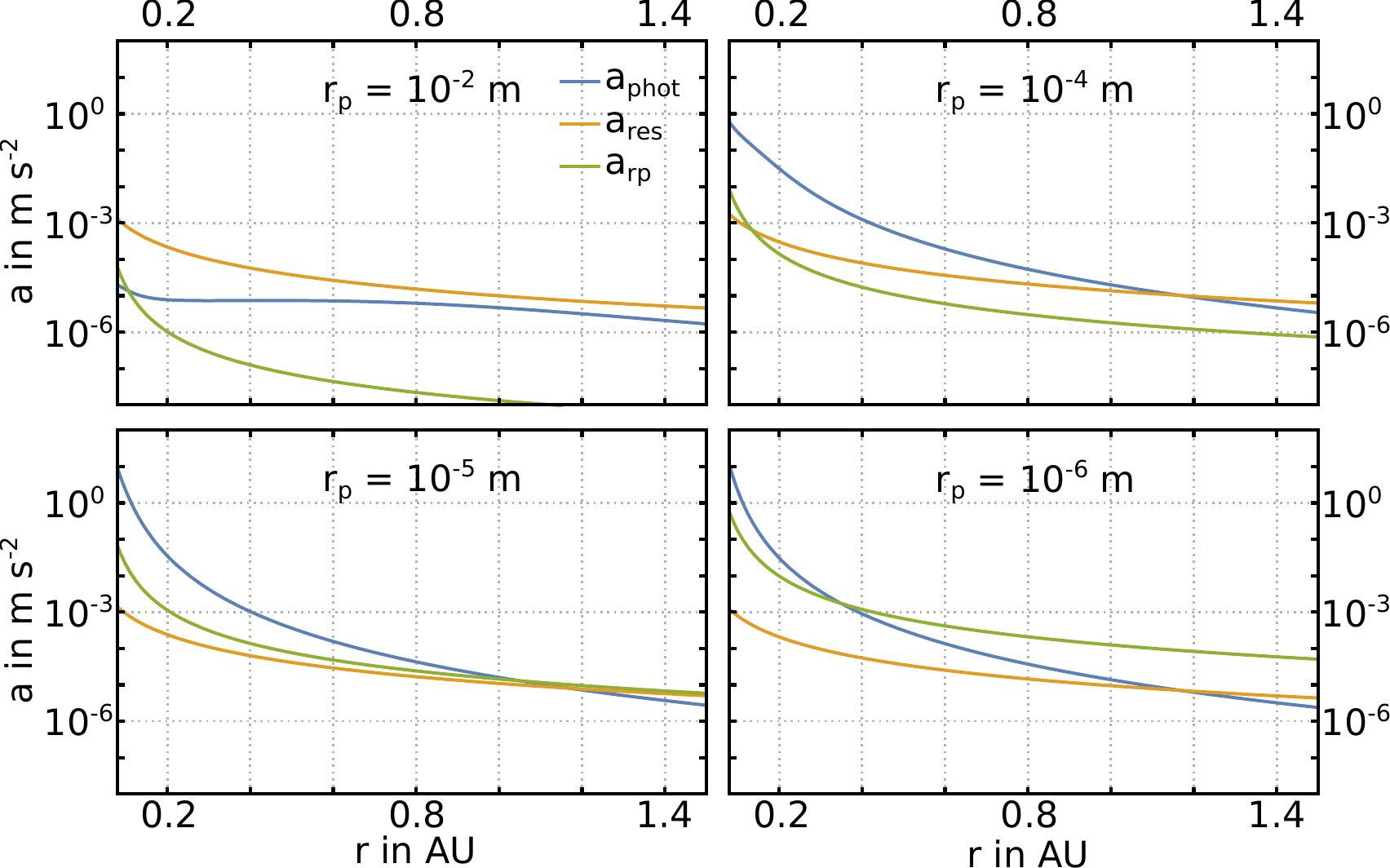}
	\caption{Absolute values of the individual accelerations of different sized particles in the disk LkCa15 (model 1). Blue denotes acceleration due to photophoresis, green due to radiation pressure and orange due to the residual gravity. Note that the acceleration due to the drag is not plotted here since it is dependent on the actual speed of the particles.}
	\label{fig:forcecompareLK}
\end{figure}

The absolute contributions of each acceleration acting on different sized particles are plotted for different distances to the central star in the disk LkCa15 in fig. \ref{fig:forcecompareLK}. 

As shown in these plots, radiation pressure is negligible for most particle sizes but dominates the force balance for smaller particles further away than a few tenth of AU. Photophoresis on the other hand can dominate the drift of particles for certain sizes and certain distances to the central star. For $10^{-2}$ m sized particles the photophoretic force is rather flat in the inner disk which results from the pressure dependency.

\begin{figure}
	\figurenum{4}
	\includegraphics[width=0.45\textwidth]{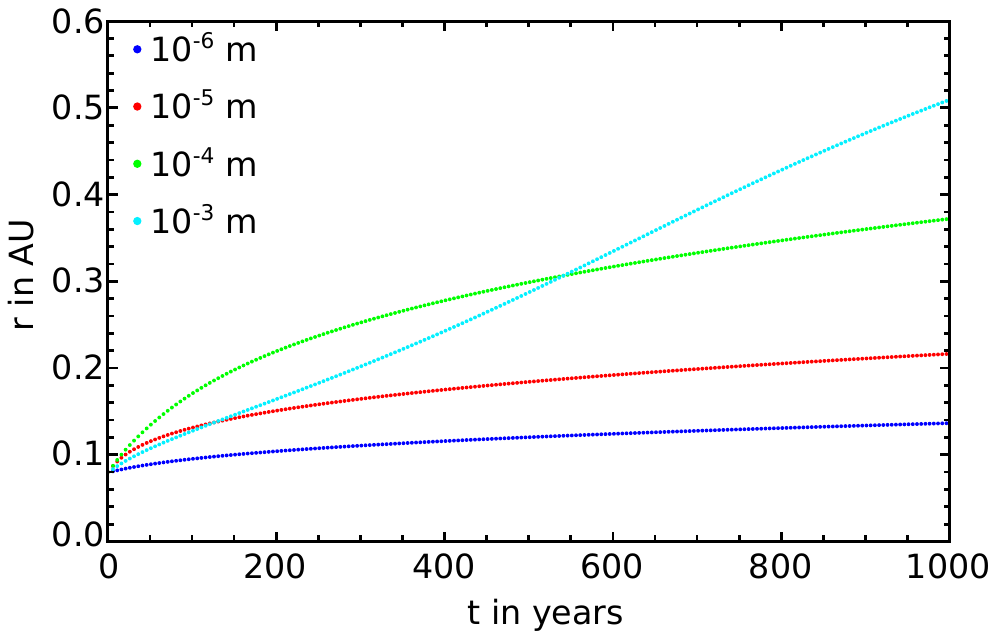}
	\caption{Radial drift for different sized particles in the disk LkCa15 (model 1).}
	\label{fig:driftLK}
\end{figure}

\begin{figure}
	\figurenum{5}
	\includegraphics[width=0.45\textwidth]{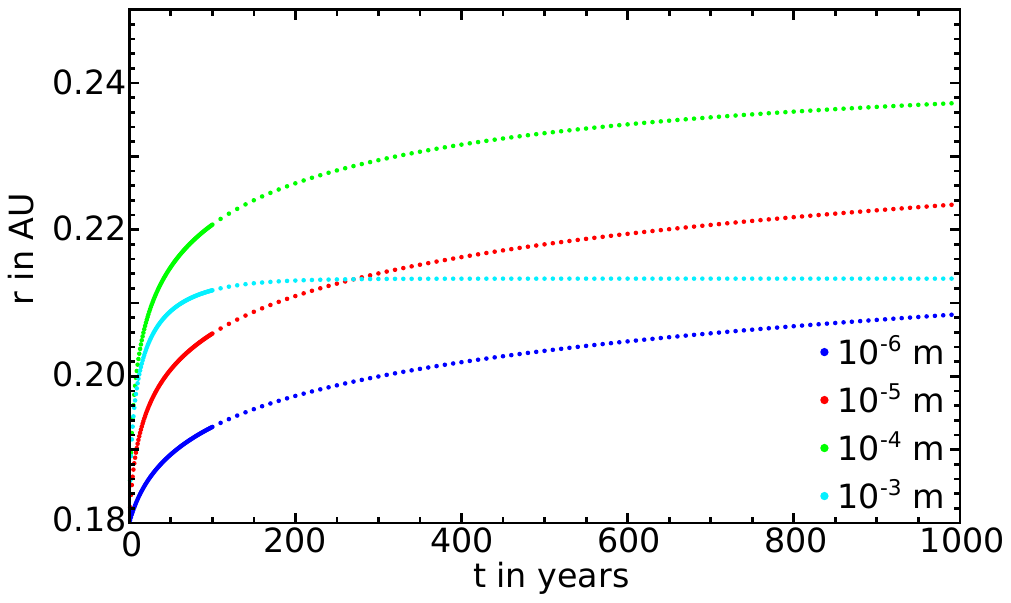}
	\caption{Radial drift for different sized particles in the disk HD135344B.}
	\label{fig:driftHD}
\end{figure}

In fig. \ref{fig:driftLK} and \ref{fig:driftHD} the radial distance $R$ over  time $t$ is plotted for four different particle sizes in the disks LkCa15 and HD135344B.


\begin{figure}
	\figurenum{6}
	\includegraphics[width=0.45\textwidth]{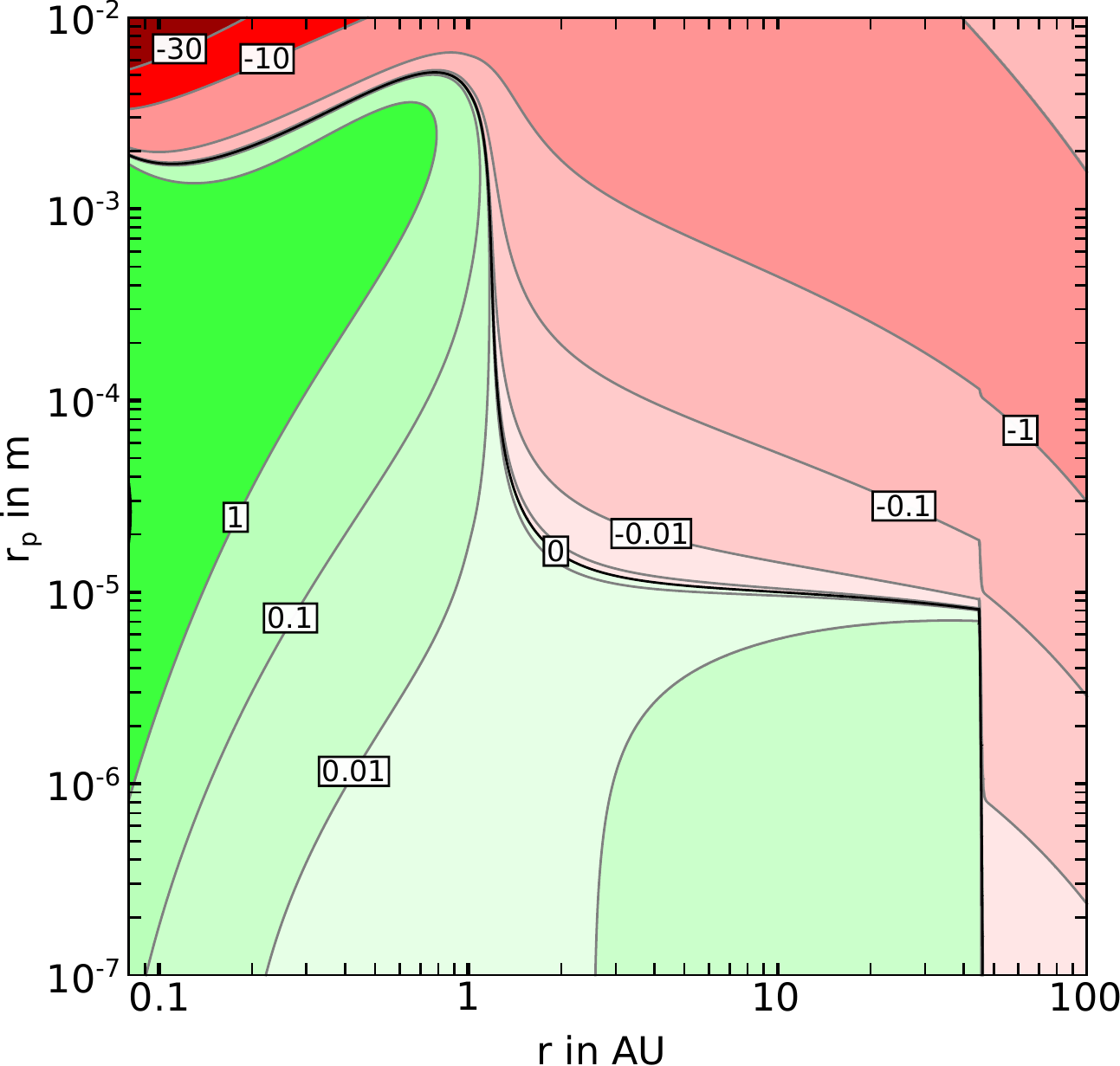}
	\caption{Radial drift for different sized particles $r_{\rm p}$ and different distances to the central star $r$ in the disk LkCa15 (model 1). Green colours denote positive values (drift away from the star), while red colours denote negative drift (towards the star). The contour velocities are in m/s.}
	\label{fig:drift3DLK}
\end{figure}

\begin{figure}
	\figurenum{7}
	\includegraphics[width=0.45\textwidth]{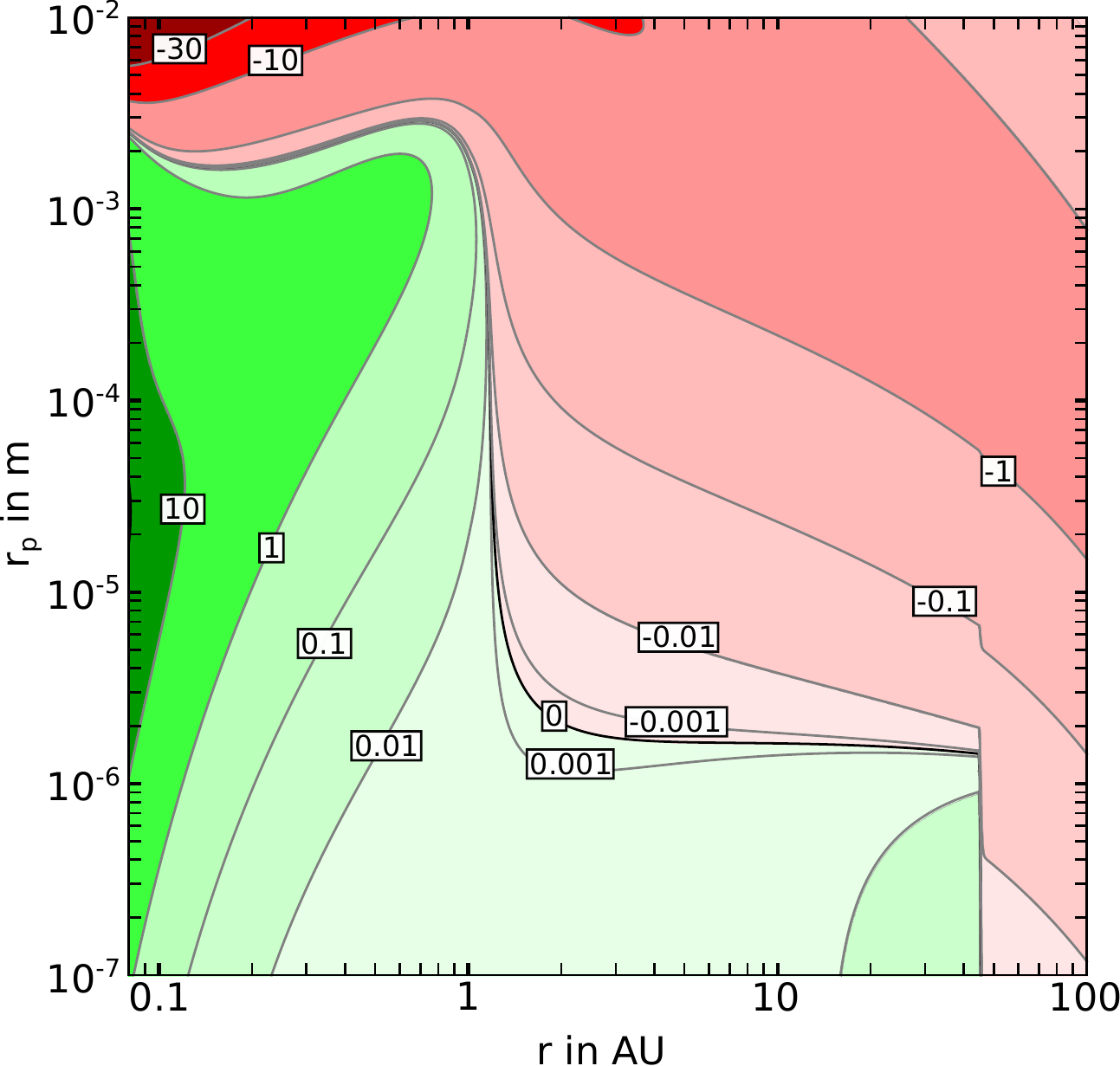}
	\caption{Radial drift for different sized particles $r_{\rm p}$ and different distances to the central star $r$ in the disk LkCa15 (model 2). Green colours denote positive values (drift away from the star), while red colours denote negative drift (towards the star). The contour velocities are in m/s.}
	\label{fig:drift3DLKM2}
\end{figure}

\begin{figure}
	\figurenum{8}
	\includegraphics[width=0.45\textwidth]{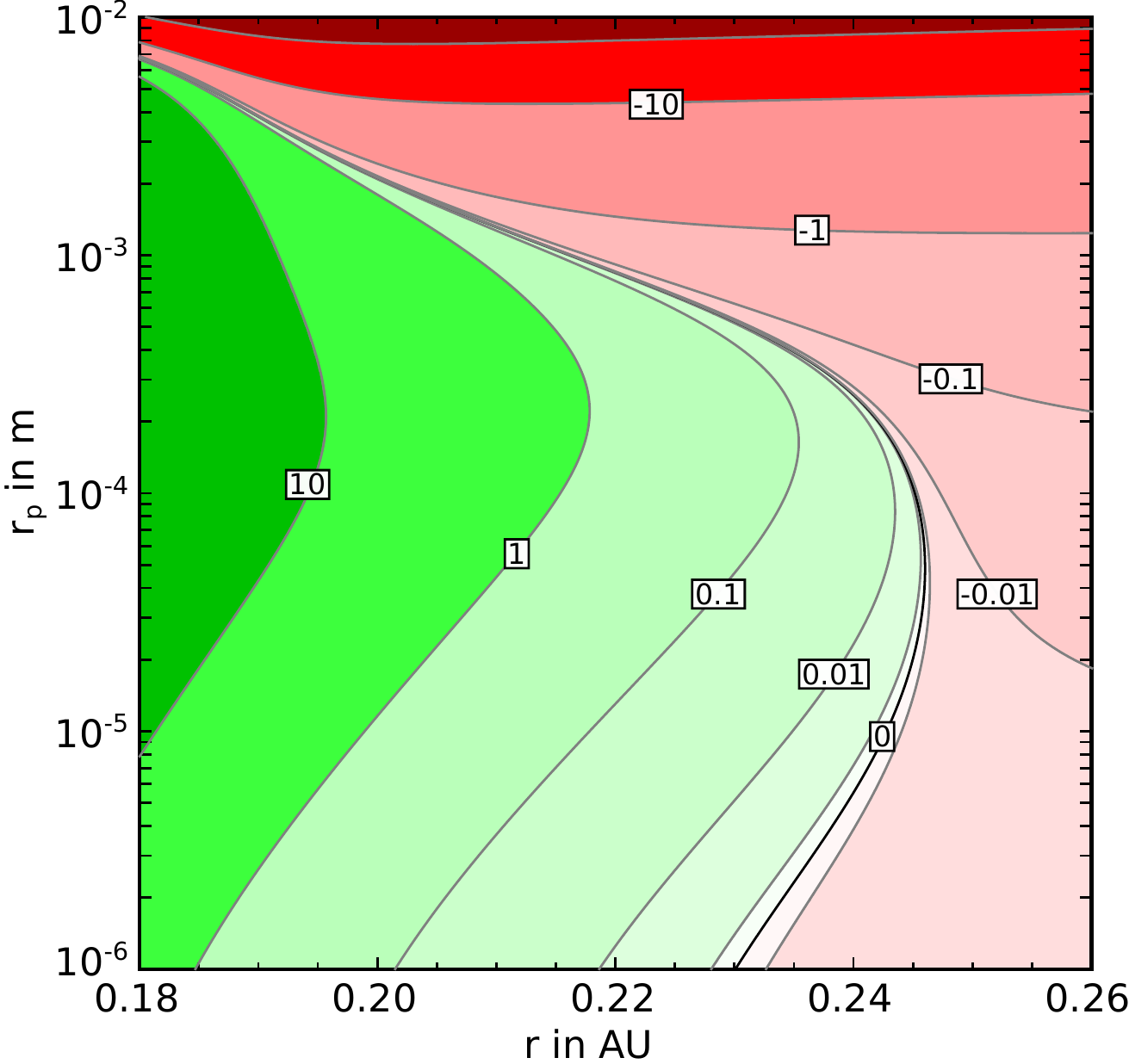}
	\caption{Radial drift for different sized particles $r_{\rm p}$ and different distances to the central star $r$ in the disk HD135344B. Green colours denote positive values (drift away from the star), while red colours denote negative drift (towards the star). The contour velocities are in m/s.}
	\label{fig:drift3DHD}
\end{figure}

Fig. \ref{fig:drift3DLK}, \ref{fig:drift3DLKM2} and \ref{fig:drift3DHD} show the calculated drift velocities of particles of different sizes $r_{\rm p}$ for different distances to the central star $r$ in the disks LkCa15 (Model 1 and Model 2) and HD135344B. Although in the original model by \cite{Andrews2011} and \cite{Marel2015} the surface density $\Sigma_\gas$ is not steady at $r_{\rm cav}$ we used the (reduced) gas surface density for the whole disk. This leads to slightly incorrect drift velocities for $r > r_{\rm cav}$ but the drift outside $r_{\rm cav}$ is of no concern here. 

Especially in the disk LkCa15, fig. \ref{fig:driftLK} shows that larger particles tend to drift faster than the small ones. Nevertheless, the exact drift behaviour is dependent on the particle size $r_{\rm p}$ and the distance to the central star $r$. Although from this plots the exact dust movement can only be estimated, it is clearly visible that particles can only drift outwards until they reach the (size specific) stopping point where the force balance is zero. As mentioned earlier and depicted in fig. \ref{fig:diskclearing}, the following mechanism is proposed:
\begin{itemize}
\item dust of different sizes is produced near $r_{\rm sub}$
\item while drifting outwards, the smallest particles grow via hit and stick to sizes around a few $10^{-6}$ m, the largest particles grow further while drifting outwards faster than the smaller ones
\item the largest particles reach their stopping point faster and can grow at this size since smaller dust is continuously drifting outwards which transfers mass onto the larger particles in collisions.
\item the growing large particles evolve along the zero-velocity line in fig. \ref{fig:drift3DLK} and \ref{fig:drift3DHD}, leading to an inwards drift. Whilst in the disk LkCa15 the large particles overcome a critical point and the drift inwards can get very high, in the disk HD135344B this evolution is slower.
\item once the large particles drift inwards they collide with other larger particles leading to erosion/fragmentation.
\end{itemize}
This recipe is detailed in the following section where single particle simulations were performed to test the model.

Fig. \ref{fig:drift3DLK} and \ref{fig:drift3DLKM2} shows that small particles can drift outwards even inside the cavity. With the cavity radius $r_{\rm cav} = 45 \, \textrm{AU}$ a critical radius $r_{\rm p, crit}$ can be determined for which the drift in the inner parts is always positive:
\begin{align}
r_{\rm p, crit}^{\rm M1} & = 2 \times 10^{-6} \, \textrm{m} \\
r_{\rm p, crit}^{\rm M2} & = 8 \times 10^{-6} \, \textrm{m} \quad .
\label{rpcrit}
\end{align}

Note here that the gas density was not modified for $r > r_{\rm cav}$ to avoid pressure bumps. The drift velocities for $r \gtrsim r_{\rm cav}$ are therefore simplified.

\section{Model for Single Particle Evolution Simulations}

Before performing single particle evolution simulations, the collision velocities and the collisional outcome has to be determined for different sized particles and different distances to the central star. In the following section we therefore describe the models for both parameters and the simulation technique. After that the results are presented and discussed.

\subsection{Collision Velocities} \label{subsec:collisionvelocities}

In general the relative velocities between particles have four different sources: radial and azimutal drift, turbulence and Brownian motion. \\

\textit{Radial and azimutal motion} - As mentioned above, we used the steady-state solution of eq. \ref{eq:radazdrift} to calculate radial and azimuthal relative velocities of the particles.

\textit{Brownian motion} - Due to the thermal motion of the gas particles, small particles are influenced by the random oriented brownian motion which results in a relative velocity dependent on the particle masses $m_1$ and $m_2$ and the distance to the central star $r$ according to
\begin{equation}
\Delta v_{\rm brown}(m_1,m_2,r) = \sqrt{\frac{8 \, k_{\rm b} \, T(r) \, (m_1 + m_2)}{\pi \, m_1 m_2} } \qquad .
\label{eq:brownianrel}
\end{equation}
Brownian motion influences collisions of the smallest particles only. 

\textit{Turbulent Motion} - We used the closed-form expressions by \cite{Ormel2007} to calculate the relative motion of particles due to turbulence in an alpha-turbulent disk \citep{Shakura1973}. The strength of turbulent velocities is yet not known since the spatial resolution of imaging only allows an estimation of an upper limit \citep{Pietu2007}. Simulations (e.g. by \cite{Dzyurkevich2010}) estimated $\alpha_{\rm turb}$ values of $10^{-3}$ in the dead zones of MRI turbulent disks. Since the inner part of a pre-transitional disk reaches only about a maximum of 1-2 AU, we consider a fully turbulent disk as shown by \cite{Desch2015}. Therefore, we picked an $\alpha_{\rm turb}$-value of $10^{-2}$. This value is consistent with estimations e.g. by \cite{Pinilla2012}, who generally assume $\alpha_{\rm turb}$ to be between $10^{-4}$ and $10^{-2}$ in the disk LkCa15 although they favor lower $\alpha_{\rm turb}$ values for the outer parts. More recent results e.g. by \cite{Hughes2011} and \cite{Guilloteau2012} show that $\alpha_{\rm turb}$ values are in the range of $10^{-2}$ to $10^{-3}$ even in the outer parts of their investigated disks (HD 163296 and TW Hya, and DW Tau respectively), but mention on the other hand that an exact calculation of the value is not possible directly from observational data and the $\alpha_{\rm turb}$ prescription might be somewhat insufficient to explain the observations in detail. Nonetheless, for the purposes here this method should be sufficient for the estimation of collision velocities due to turbulent motion and the high $\alpha_{\rm turb}$ values are favored which are also predicted by \cite{Chiang2007} at the edge of transitional disks.
For the disk HD135344B we picked $\alpha_{\rm turb} = 3 \times 10^{-2}$.

\begin{figure}
	\figurenum{9}
	\includegraphics[width=0.45\textwidth]{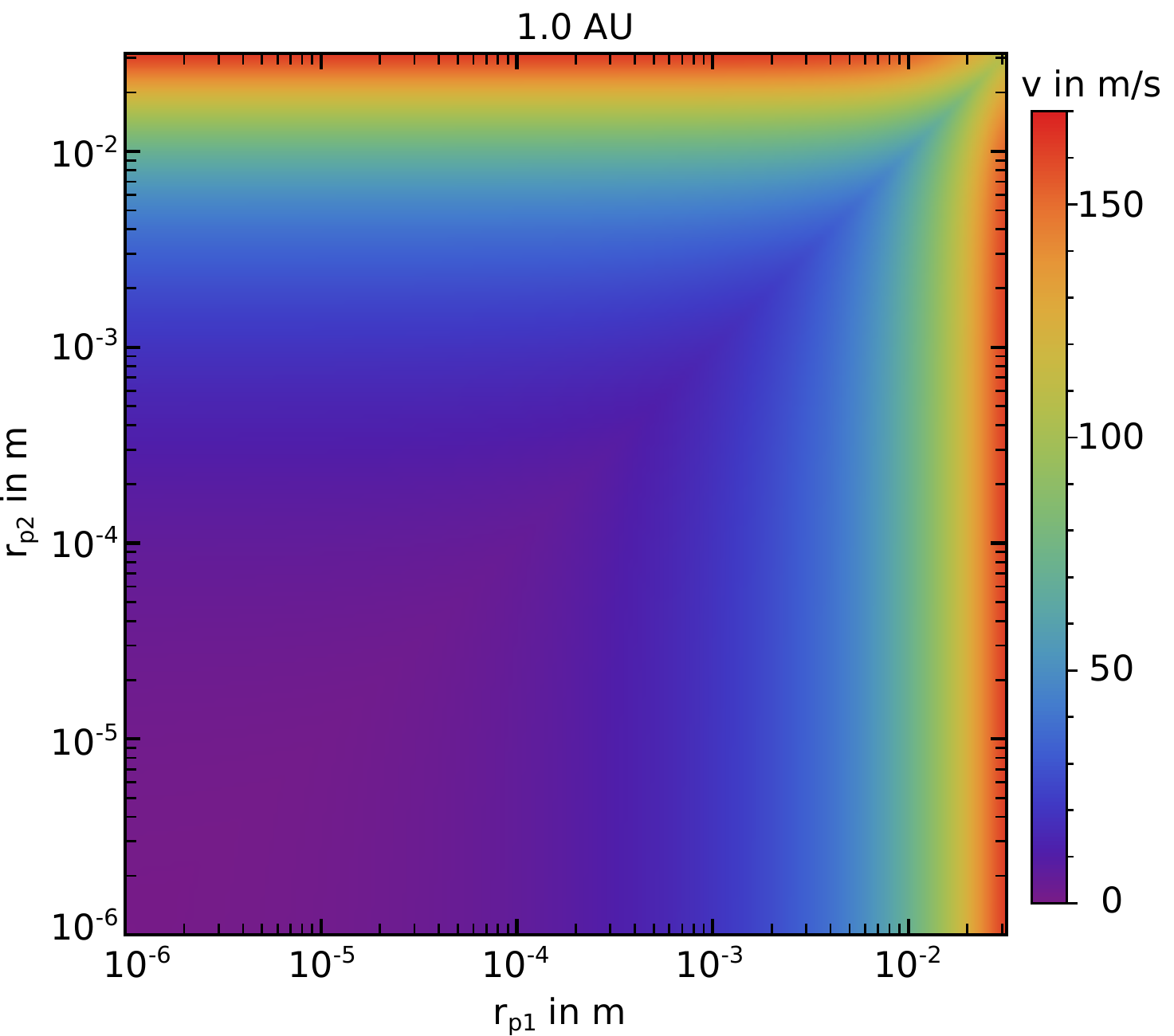}
\vspace{2 mm}
	\includegraphics[width=0.45\textwidth]{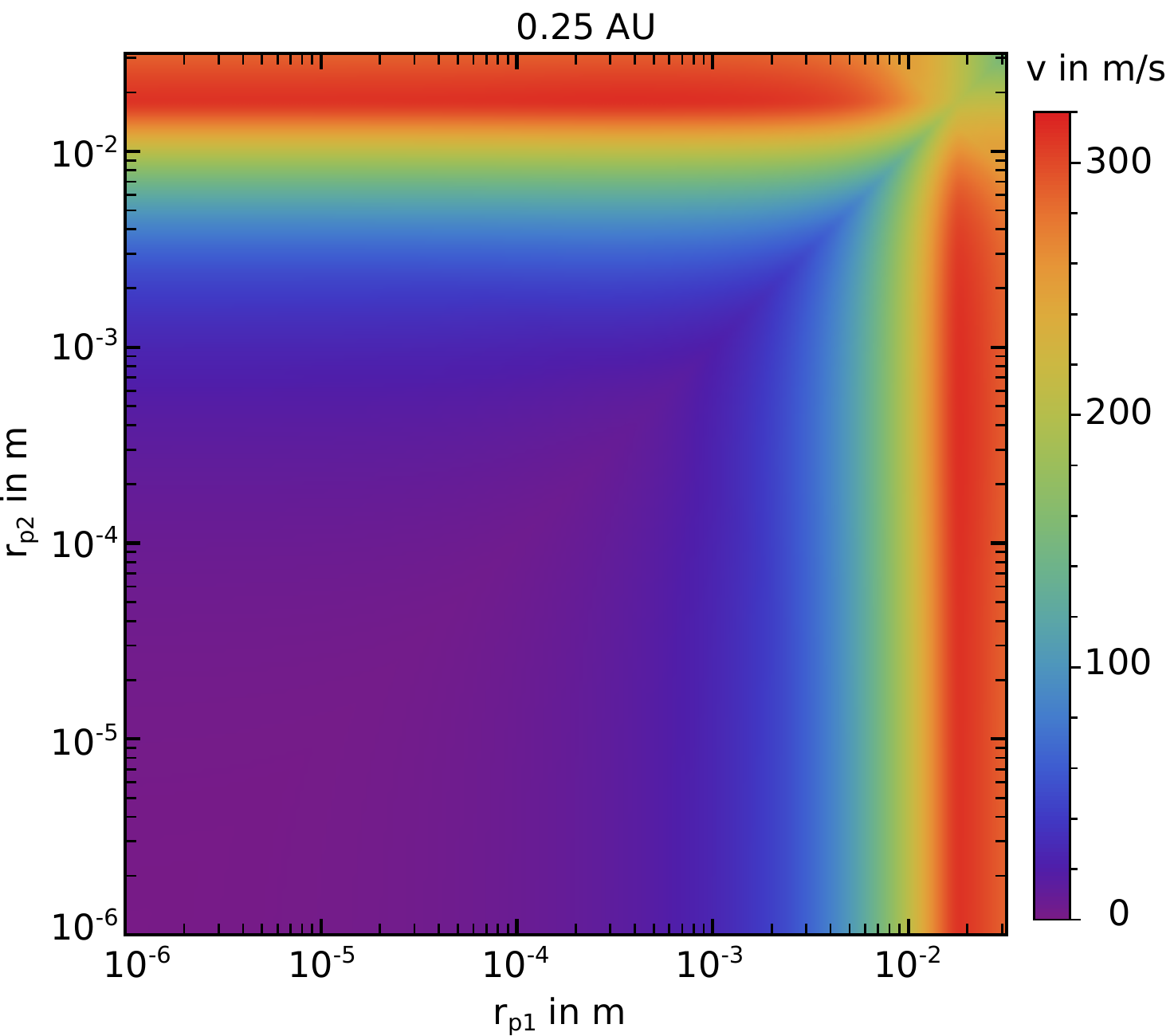}
	\caption{Collision velocities for different sized particles in the disk LkCa15 (model 1) at 1.0 AU (top) and 0.25 AU (bottom) for $\alpha_{\rm turb} = 10^{-2}$.}
	\label{fig:collisionvelocitiesLK}
\end{figure}

In fig. \ref{fig:collisionvelocitiesLK} the relative velocities in the disk LkCa15 (model 1) for 1 AU and 0.25 AUs are plotted for $\alpha_{\rm turb} = 10^{-2}$. Note here that the collision velocities can reach values above 300 m/s close to the star.

\subsection{Collisional Outcomes}

We used the model by \cite{Windmark2012a} to calculate the collisional outcome of particles at different distances to the central star. In their model collisions between particles can result in basically five different outcomes:
\begin{enumerate}
 \item sticking - small particles can stick directly to each other if the collision velocities are not too high
 \item bouncing - for increasing collision velocities or particle sizes bouncing can occur
 \item fragmentation - if equal sized particles collide with very high velocities, catastrophic disruption can occur
 \item mass transfer - if different sized particles collide at lower velocities mass gain of the target dominates over erosion resulting in net mass transfer
 \item erosion - for increasing collision velocities or particle sizes erosion starts to dominate over mass gain, resulting in net mass loss
\end{enumerate}
\textit{Sticking and Bouncing} - According to \cite{Weidling2012} the mass-dependent sticking and bouncing velocities are dependent on the smaller particle mass $m_{\rm small}$ as follows
\begin{align}
\Delta v_{\rm stick}(m_{\rm small}) & = 10^{-2} \, \left( \frac{m_{\rm small}}{m_{\rm stick}} \right)^{-5/18} \frac{\rm m}{\rm s} \\
\Delta v_{\rm bounce}(m_{\rm small}) & = 10^{-2} \, \left( \frac{m_{\rm bounce}}{m_{\rm stick}} \right)^{-5/18} \frac{\rm m}{\rm s}
\label{eq:vstickbounce}
\end{align}
with $m_{\rm stick} = 3.0 \times 10^{-15}$ kg and $m_{\rm bounce} = 3.3 \times 10^{-6}$ kg. If collision velocities are below $v_{\rm stick}$, particles stick together in collisions, if collision velocities are above $v_{\rm bounce}$, particles always bounce or - if the collision velocity is too high - mass transfer/erosion/fragmentation occurs. For collision velocities between $v_{\rm stick}$ and $v_{\rm bounce}$, the both events can occur with a specific sticking probability. \\
\textit{Fragmentation} - Complete destruction of both collision partners occurs only for (more or less) equal sized, larger particles (in the disk LkCa15 this size has to be above approximately $4 \times 10^{-5}$ m for the velocities shown in \ref{fig:collisionvelocitiesLK}). For our estimations this plays a role in the recycling process later on when larger bodies begin to drift towards the star. \\
\textit{Mass Transfer and Erosion} - The most important regime in our estimations is the mass transfer regime, where the smaller collision partner is destroyed during the collision and mass is transferred from the smaller to the larger particle. While \cite{Windmark2012a} described a possible runaway growth of single, "lucky" particles, for us this regime is beneficial for the sweep-up process with the help of larger particles. The amount of mass transferred during such a collision can be described via the difference of transferred and eroded mass,
\begin{equation}
m_{\rm gained} = m_{\rm transferred} - m_{\rm eroded} = p_{\rm mt} \cdot m_{\rm impactor} \qquad .
\label{eq:massgained}
\end{equation}
Once the eroded mass is larger then the transferred one, the larger body is loosing mass as well. For details on the calculation of the transferred and eroded mass see \cite{Windmark2012a} for further details. 

\begin{figure}
	\figurenum{10}
	\includegraphics[width=0.45\textwidth]{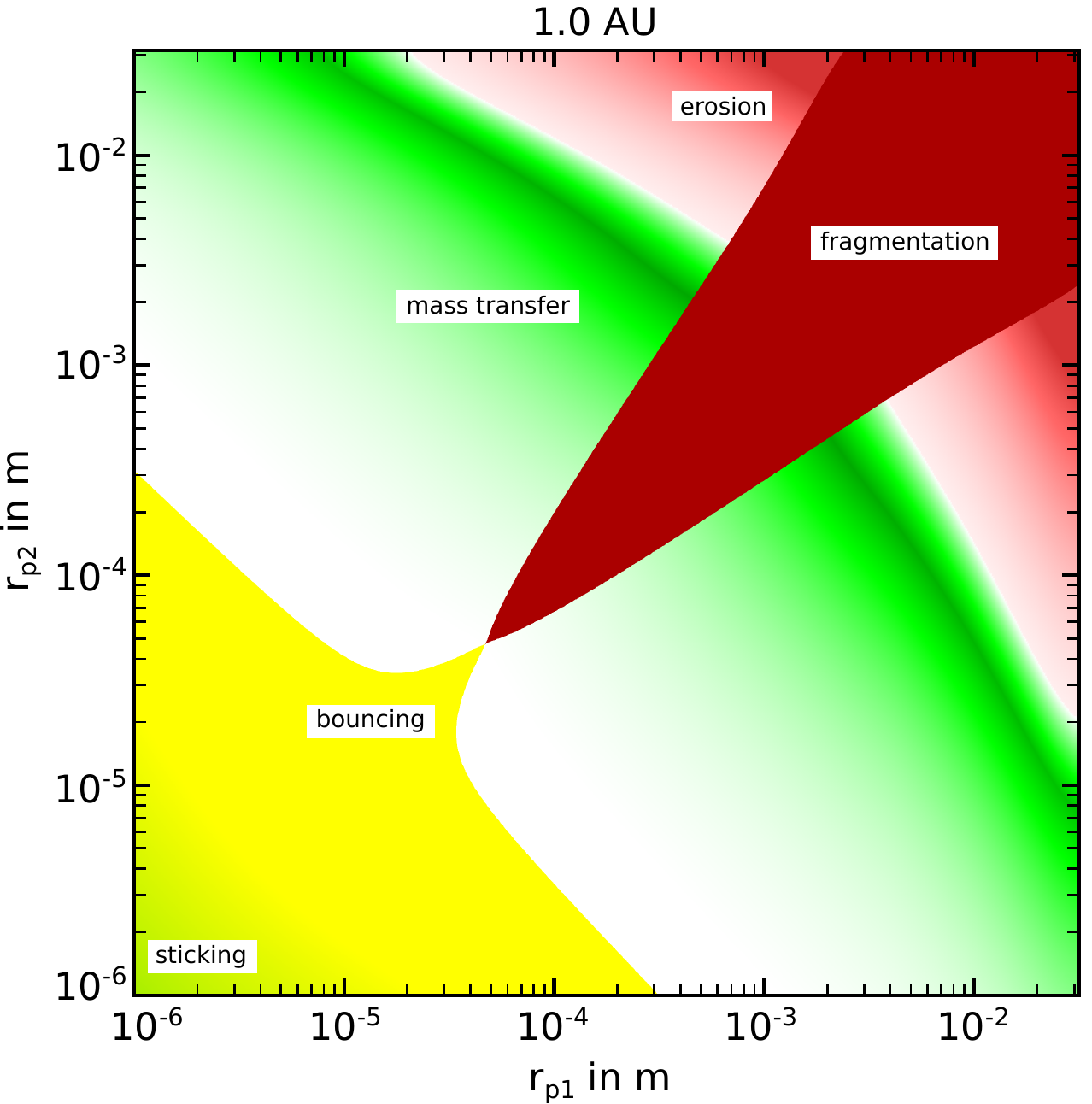}
\vspace{2 mm}
	\includegraphics[width=0.45\textwidth]{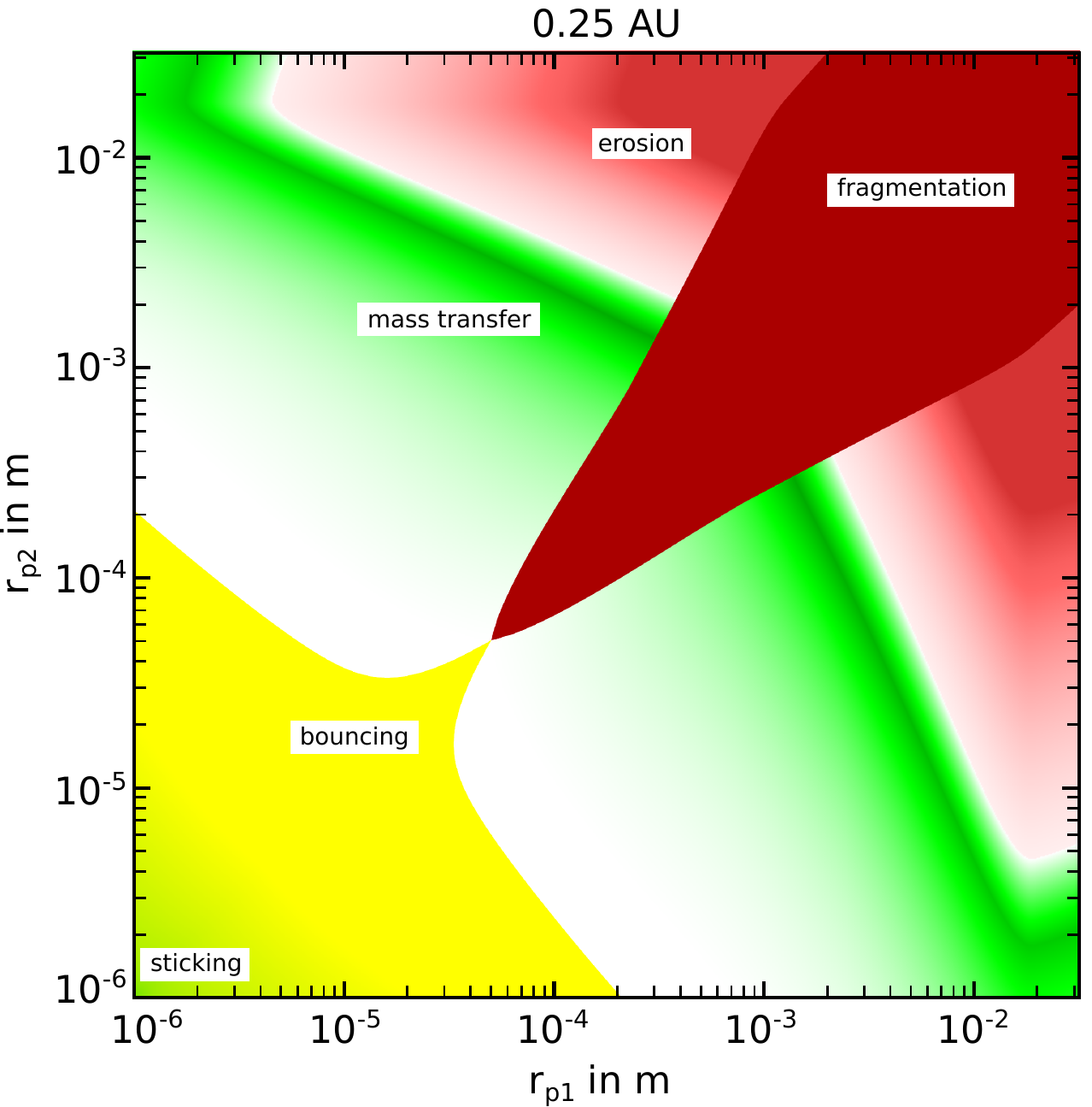}
\caption{Collisional outcome of different sized particles in the disk LkCa15 (model 1) at 1 AU (top) and 0.25 AU (bottom); green denotes growth (direct sticking or mass transfer), yellow denotes no change in mass and red denotes mass loss (erosion or fragmentation). The transferred mass is color coded here from white (no mass gain) to dark green (about 0.16 $m_{\rm impactor}$). The eroded mass is color coded from white (no mass loss) to red (mass loss up to 4 $m_{\rm impactor}$).}
  \label{fig:colOutcome}
\end{figure}

Examples of collisional outcome in the disk LkCa15 (model 1) for different particle sizes 0.25 AU and 1.0 AU are shown in fig. \ref{fig:colOutcome}. 

\subsection{Sizes of Produced Dust}

Sweeping pebbles have to be produced between the sublimation radius $r_{\rm sub}$ and the gap radius $r_{\rm gap}$. We consider destructive aggregate collisions as source for these particles, while the size of these aggregates are on the order of $10^{-3}$ m to $10^{-1}$ m. We assume a power law dependence for the radii distribution (like e.g. \cite{Deckers2014}, \cite{MacGregor2016}):

\begin{equation}
n(r_{\rm p}) = a_{\rm n} \cdot r_{\rm p}^{-\kappa_{\rm size}} \qquad .
\label{eq:sizedistribution}
\end{equation}

Depending on $\kappa_{\rm size}$, one has to assume a cutoff for the maximum and/or minimum mass produced (called $r_{\rm p, max}$ and $r_{\rm p, min}$). For our models the values/functions are listed in tab. \ref{table:diskparameters2}. For the disk LkCa15, we assumed $\kappa_{\rm size}$ to be dependent on the distance to the central star $r$ via a function
\begin{equation}
\kappa_{\rm size} = 3.1 + (2 \, \frac{r - r_{\rm sub}}{\rm AU})^{0.4}	\quad .
\label{eq:kappasizeLK}
\end{equation}
The radial dependency should satisfy the general model of recycling. Near the sublimation radius, small particles can only be created by catastrophic disruption of large bodies. Therefore, the size distribution should be equaling the collisional outcome distribution as depicted e.g. by \cite{Deckers2014}. On the other hand, further outwards the concentration of small particles should be increasing since large bodies sweep up only parts of the smaller ones, resulting in disruption of the small bodies and producing smaller particles. Furthermore, small particles who drift outwards are slowed down as can be seen in fig. \ref{fig:drift3DLK} and \ref{fig:drift3DLKM2}. This leads to an enrichment of small particles in the outer parts. For the disk HD135344B, $\kappa_{\rm size}$ was chosen to be constant since the inner dust disk is much smaller and a recycling process does not necessarily happen throughout the whole disk but might happen on local scales. The value of $\kappa_{\rm size} = 4.0$ was chosen since multiple destructive events can lead to a significant enrichment of small particles and therefore to higher $\kappa_{\rm size}$ values. Note here that in general $\kappa_{\rm size}$ is a highly uncertain parameter and depending on the exact size- and spacial evolution of the dust distribution.

\subsection{Simulation Details}

The simulations were performed as Monte Carlo simulation, following the subsequent procedure:

\begin{enumerate}
\item insert particle of size $r_{\rm p}$ at distance to the central star $r$
\item calculate drifted distance for time interval $\Delta_{\rm t}$
\item calculate collision probabilities for particle within size-bin $i$ following $p(r_{\rm p}, i) = N(i) \Delta v_{r_{\rm p},r_{\rm p}(i)} \sigma_{r_{\rm p},r_{\rm p}(i)}$ with $\sigma_{r_{\rm p},r_{\rm p}(i)}$ is the cross-section for a collision between two particles of size $r_{\rm p}$ and $r_{\rm p}(i)$
\item if collision with fragmentation occurs, calculate new particle mass/size (largest fragment)
\item for non-fragmenting collisions calculate mass change in time interval $\Delta_{\rm t}$ and therefore new particle mass/size
\end{enumerate}

Steps 2-5 are repeated and the particle evolution ($r_{\rm p}(t)$, $r(t)$) is investigated. For the calculations we binned the size distribution (bins per size decade) with a resolution of 30.

\section{Results for single particle evolution simulations}

Until  otherwise stated, we used the parameters from tab. \ref{table:diskparameters} with some additional parameters which can be found in tab. \ref{table:diskparameters2}.  
From the drift behavior (see fig. \ref{fig:drift3DLK}) we would assume that the dust density $\rho_\dust$ would change during the disks evolution and more particles would be concentrated at the outer part of the inner dust ring. 
To account for this, we varied $\kappa_{\rm size}$ with distance to the central star $r$.

\begin{table}
\caption{Simulation/Disk parameters}              
\label{table:diskparameters2}      
\centering                                      
\begin{tabular}{l l l l l l l l l l}          
\hline\hline                        
Disk 				&	$\alpha_{\rm turb}$ 	&	$\kappa_{\rm size}$									& $r_{\rm p, min}$   	& $r_{\rm p, max}$	\\
				&					&													& (m)			& (m)			\\
\hline
LkCa15 Model 1		&	$10^{-2}$			&	$3.1 + \left(2 \frac{r - r_{\rm sub}}{\textrm{AU}} \right)^{0.4}$		&  $10^{-6}$		& $10^{-2}$		\\
LkCa15 Model 2		&	$10^{-2}$			&	$3.1 + \left(2 \frac{r - r_{\rm sub}}{\textrm{AU}} \right)^{0.4}$		&  $10^{-6}$		& $10^{-2}$		\\
HD134355B		&	$3 \times 10^{-2}$			&	$4.0$		&  dyn.		& dyn.		\\
\hline                                   
\hline
\end{tabular}
\end{table}

\subsection{LkCa15}

\begin{figure}
	\figurenum{11}
	\centering
	\includegraphics[width=0.45\textwidth]{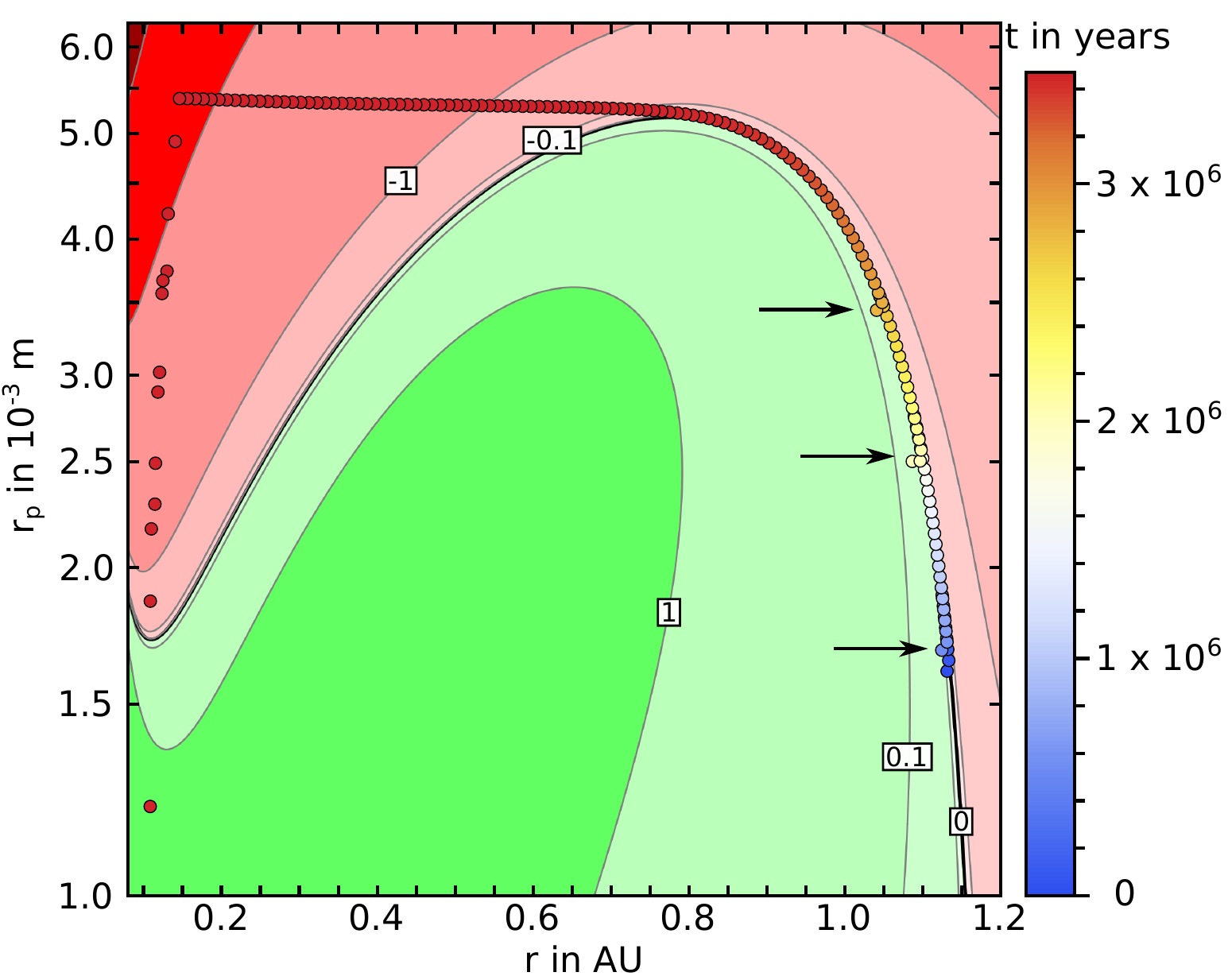}
\vspace{2 mm}
	\centering
	\includegraphics[width=0.45\textwidth]{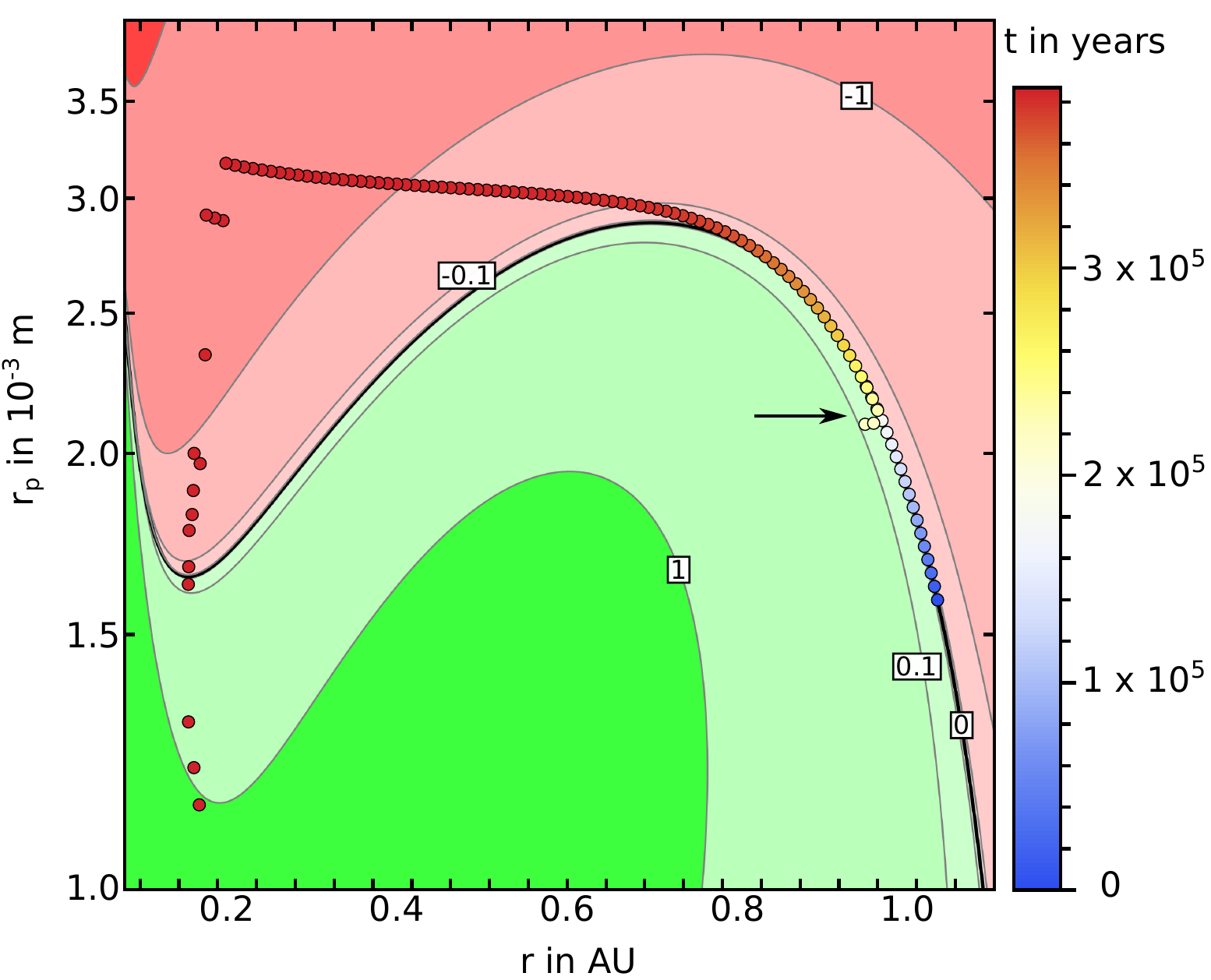}
	\caption{Example of simulated particle drifts for LkCa15; top: model 1, bottom: model 2; The trajectory is color coded according to time. Arrows mark distructive events or erosion.	}
	\label{fig:growthtrajectoriesLK}
\end{figure}

Two example growth trajectories are plottet in fig. \ref{fig:growthtrajectoriesLK} for the first and the second model respectively. Note here that for visualization not every data point is plotted but only a selection. As starting point, the maximum size of a particle coming from the sublimation radius was chosen (in both cases $1.6 \times 10^{-3}$ m). With this assumption, no growth/destruction had occurred during the outwards drift. The maximum size where drift is still positive is much smaller in the second model. Therefore, particles do not have to grow as much as in the first model and can be routed back into the recycling process much faster. This can be clearly seen in the different times necessary for growth. In the first model the timescale is on the order of $10^6$ years, while in the second model this value is around $10^5$ years. Also in model 2 larger particles show positive drift up to $2.5 \times 10^{-3}$ m near the sublimation radius. This leads to a steady state of larger particles which can help destroying the inward drifting larger particles which had overcome the bump at larger distances to the central star. 
Analyzing the trajectories from fig. \ref{fig:growthtrajectoriesLK} in detail, destructive events occur even at the zero-velocity line (marked by the black arrows). In the lower plot, the particle is loosing $\sim 20 \%$ of its mass leading to a growth delay of $\sim 4 \times 10^4$ years. Since the simulation timescale $\Delta t$ is dynamically adopted to the radial velocity of the particles and generally small, the probabilities for disruptive events in the interval $\Delta t$ are low as well. Since destructive events get more probable for smaller grains, the grown particles can drift inwards quite far until the first fragmentation occurs and a disintegration cascade sets in.

\subsection{HD135344B}

\begin{figure}
	\figurenum{12}
	\centering
	\includegraphics[width=0.45\textwidth]{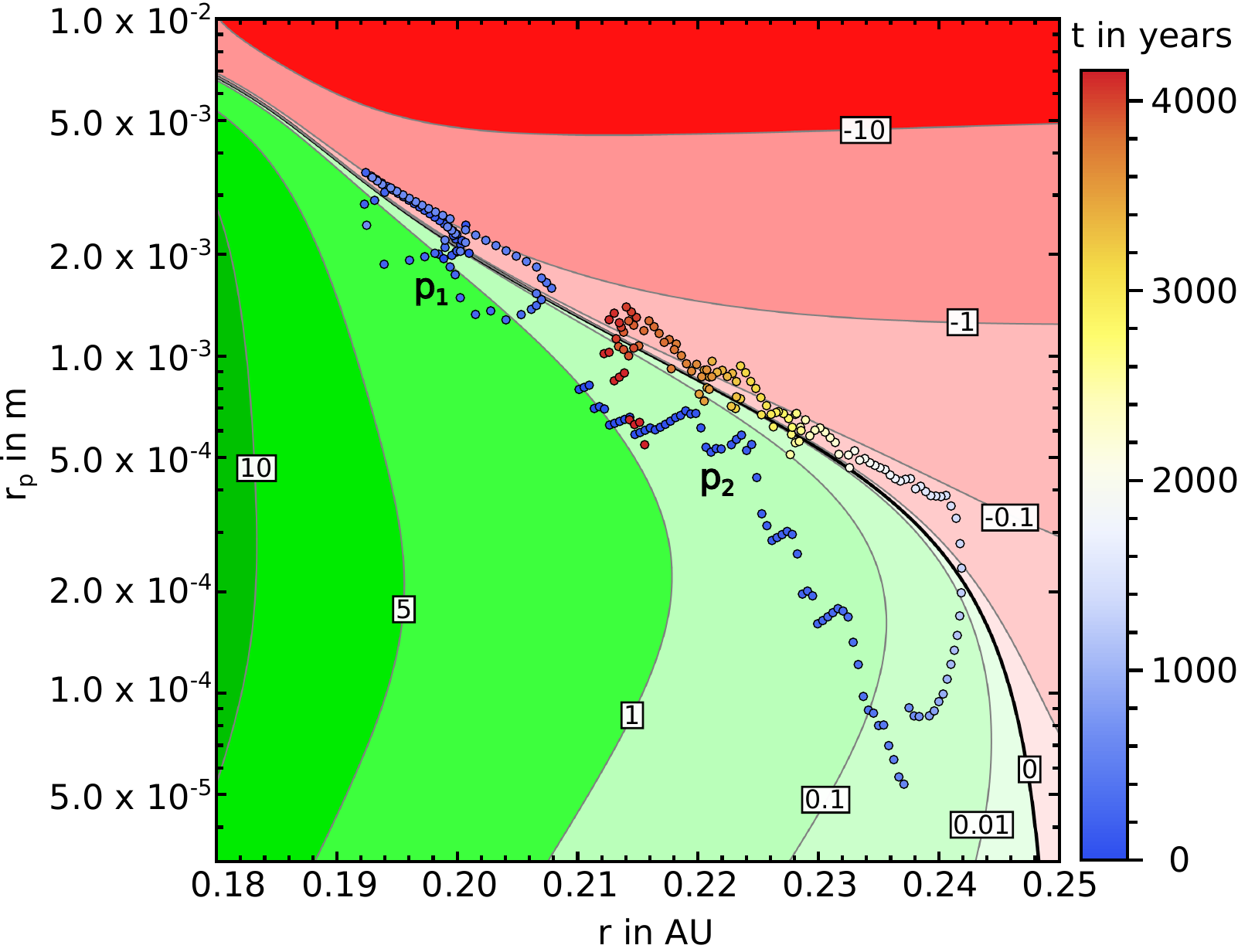}
	\caption{Example of two simulated particle drifts (p$_1$ and p$_2$) for the disk HD135344B. The trajectories are color coded according to the time. For visibility reasons, Particle p$_1$ is plotted only for the first 1000 years while particle p$_2$ is plotted for the full timescale displayed.Due to the high amount of dust, the particles suffer from destructive events much more often than in the disk LkCa15. As can be seen here, particles therefore do often not drift the complete disk but only small parts of it.}
	\label{fig:growthtrajectoriesHD}
\end{figure}

Two different growth trajectories are plotted in fig. \ref{fig:growthtrajectoriesHD}. 
To account for the lack of very small particles further outwards, we adopt the particle size limits dynamically according the maximum and minimum particle size present at a distance $r$ to the central star. This was done by using the size- and distance dependent radial drift velocity (see fig. \ref{fig:drift3DHD}) and determining the maximum and minimum particle size via the zero-velocity line. For distances $r$ smaller than $\sim 0.23$ AU, the minimum particle size was set to $10^{-6}$ m. Via this method, the minimum and maximum particle sizes at 0.24 AU for example are about $8 \times 10^{-6}$ m and $3 \times 10^{-4}$ m respectively. Note here that by applying these limits the creation of small particles due to collisions in the outer disk is neglected.

\section{DISCUSSION AND CAVEATS}

Generally, the velocity calculations show that the inner dust disk in pre-transitional disks can be in a steady state with only little mass lost due to either outwards drift if particles are very small or sublimation of inwards drifting larger aggregates. Particles drifting further outwards can explain observations of (very little) dust inside the cavity. Nonetheless, since we use approximations for the photophoretic force which are generally valid for optically thin disks where the background radiation correspond to a fixed temperature, the drift velocities might differ in both directions: inside the inner disk the temperature gradient on the particle (and therefore the photophoretic force) might be overestimated due to not including infrared radiation coming from both radial directions, while on the outer edge of the inner dust ring close to $r_{\rm gap}$ the temperature gradient might be underestimated since the infrared radiation coming from the inner disk is not included in the model while the stellar radiation is already tenuated. Therefore the calculated disk edges are not strict but might be diffuse as well while the drift timescale might be underestimated. Including these aspects would need improvement in  modeling the temperature profiles of illuminated, complex particles. 
The Monte Carlo simulations are based upon a collision model, modification of this model would lead to changing growth timescales. 
Without loss of generality though, the approximation used describes the general particle movement.
\subsection{LkCa15}

\textit{Inner Disk} -- As shown in fig \ref{fig:drift3DLK} and \ref{fig:drift3DLKM2} as well as in the single particle evolution simulations (fig. \ref{fig:growthtrajectoriesLK}), a self sustained recycling process can be established in the disk with small particles drifting outwards while some of them grow beyond the turning point (the maximum size where drift can still be positive) and fall back inwards where they get destroyed. Parameter sweeps for the disk with varying $k_{\rm th}$, $\chi_\dust$, $\delta_\gas$, $\delta_\dust$, and $\kappa_{\rm med}$ show that several configurations exist with self-sustained recycling (see appendix tab. \ref{table:parametersweep} for some examples). So the process does not need extreme fine tuning. Nonetheless, we note that the growth timescales can differ significantly as shown in the comparison between the two example models. \\
Beneficial for a self-sustained recycling process in a steady state disk are growth timescales similar to the drift timescales for the particles which include the most mass. Since generally $\kappa_{\rm size}$ is assumed to be greater than 3 (a discussion on this was e.g. held by \cite{MacGregor2016}), most mass is located in the smallest particles. These particles drift with velocities $\sim 0.1 - 0.005$ m/s or 1 AU every $\sim 50\,000 - 1\,000\,000$ years. 

\begin{figure}[ht]
	\figurenum{13}
	\centering
	\includegraphics[width=0.45\textwidth]{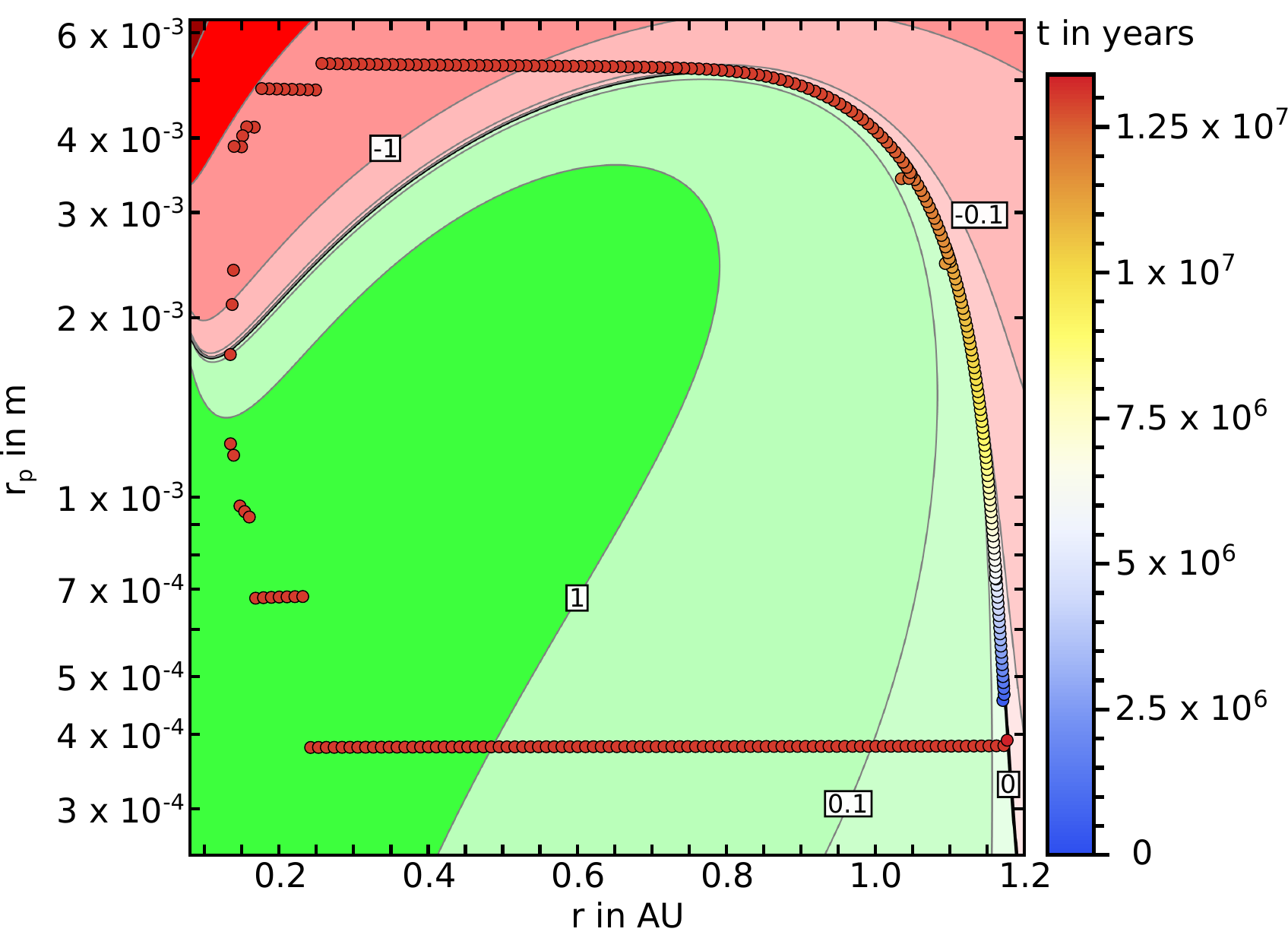}
	\caption{Example of simulated particle drifts for the disk LkCa15, model 1. The contours mark the drift velocities, positive (green) is outwards drift, negative (red) inwards drift. The trajectory is color coded according to the time. A full trajectory is shown with a total time of 10 Myr.}
	\label{fig:growthtrajectoriesLKFULL}
\end{figure}

Fig. \ref{fig:growthtrajectoriesLKFULL} shows a full example trajectory for model 1. As shown there, the total time needed can exceed $10^7$ years if the growing particle is small. Since the turning point in the model 2 is at smaller particle sizes, a full trajectory can take significantly shorter than in model 1. Nonetheless, the simulations show that the drift timescales and growth timescales are generally on the same order and a steady disk state could be reached including the dynamic of the stated recycling process. Even if drift and growth timescales would differ, a self sustained recycling process could be established: assuming that the growth is not quick enough, this leads to more mass concentrating on the outer edge of the inner dust ring which would decrease the growth timescale leading to more mass being transported inwards again. This would lead to a depletion of mass on the outer rim and to more mass near $r_{\rm sub}$. The dust surface density $\Sigma_\dust$ would therefore oscillate in time with different phases for different distances to the central star. \\
The general model for the disks surface density and temperature profile is somewhat simplified and might be differing from the real disk properties. Detailed studies on the temperature distribution (and therefore on the disks scale height) might be necessary using heat transfer modeling. Furthermore, an even more detailed model would include dust movement not only in radial but also in vertical direction. Nonetheless, we assume that the general picture is not changing since several disk parameters exist which influence the motion of particles heavily (dust scale height, surface densities, dust drop values, e.t.c.). Therefore multiple parameter constellations might exist in this case as well where a recycling process will be stable. Changes in the height profile would only influence the gap radius position since the pressure ratio $p / p_{\rm opt}$ and therefore the strength of the photophoretic force will be changed. Note here that the outwards motion of particles and the disk height would be coupled when calculating the correct temperature profile via radiative heat transfer modeling. Nonetheless, further analysis have to be made and we dedicate this to future work. 
Comparing the drift and growth timescales to the viscous timescale at 1.0 AU for Model 2
\begin{equation}
t_{\nu, \rm turb} = \frac{r^2}{\alpha_{\rm turb} \, c_{\rm s} \, H} \approx 10^5 \, \mathrm{yr}
\label{eq:viscoustimescale}
\end{equation}
,shows that the recycling process is independent on the gas behavior (note here that the gas drift is already included in the calculations) and might be present as long as gas is replenished from inside the cavity. Once this gas supply is lowered / cut -- for example due to growth of a large planet -- the recycling process might collapse and the dust will begin to vanish either by inwards drift (larger particles) or by outwards drift due to radiation pressure (smaller particles). Assuming this evolution, it might explain why there exist pre-transitional disks with large gas- and dust free inner disk parts up to several AU which can be seen for example in the disk SR21 \citep{Marel2015}. As depicted there, the disk is free of gas and dust from the sublimation radius to 7 AU and only little gas and dust from 7 AU to the inner radius of the outer dust disk which is at about 25 AU.

\textit{Outer Disk} -- The drift calculations for both disk models (fig. \ref{fig:drift3DLK}, \ref{fig:drift3DLKM2}) show positive drift velocities inside the cavity for particles smaller than $r_{\rm p, crit}^{\rm M1} = 2 \times 10^{-6} \, \textrm{m}$ or
$r_{\rm p, crit}^{\rm M2} = 8 \times 10^{-6} \, \textrm{m}$ respectively. Therefore the cavity might be cleaned from small dust by radiation pressure. Although seemingly in contrast to work by \cite{Dominik2011} this is not so. First of all, they used other disk models. Secondly, we only present calculated drifts for single particles based on an opacity model for the inner disk. The dynamics of multiple grains (pile up) is not considered here in detail. While the force balance is positive for single particles, dust pile up, the resulting reduction of the optical depth as well as viscous mixing might influence the results heavily. \cite{Dominik2011} therefore mentioned that though drift velocities can be positive, the mass flux at $r_{\rm cav}$ points towards the star for their investigated disks. However, the processes at the outer edge are beyond the scope of this work.

\subsection{HD135344B}

HD135344B is significantly different from LkCa15. The disk is optically much thicker than LkCa15 with an estimated inner disk size of 0.07 AU (from 0.18 AU to 0.25 AU) (e.g. \cite{Marel2015}) and only little radiation can reach the outer edge of the inner disk. Due to the gas accretion, small particles ($\sim 10^{-6}$ m) are unable to drift as far outwards. Particles of $\sim 5 \times 10^{-5}$ m drift the furthest and essentially just reach $r_{gap}$. 
This leads to more locally arranged recycling mechanisms taking place all over the disk with many possible particle trajectories (fig. \ref{fig:growthtrajectoriesHD}). At $0.25$ AU where $10^{-4}$ m particles reach the outermost point and can either get destroyed in collisions with similar sized particles and drift inwards or grow and drift inwards as well. This feature prevents the smallest particles from being trapped further outwards since these particles are very unlikely to overcome the bouncing barrier directly without help by a larger particle (see the ideas by \cite{Windmark2012a} and \cite{Windmark2012}). Therefore, particles observed in the cavity cannot have been released from the inner disk.

\subsection{Impact of turbulence on the recycling mechanism}
Since $\alpha_{\rm turb}$ assumed in the disks is highly uncertain, the influence of differing values on the collisional outcome and the growth is discussed in this section. In general, our model benefits from large $\alpha_{\rm turb}$ values. This is caused by the fact that large values imply larger collision velocities and therefore a shift of the transition between bouncing and mass transfer towards smaller particle sizes (see the collisional outcome plots in fig. \ref{fig:colOutcome}). Furthermore, higher collision velocities benefit particle growth due to mass transfer in the outer parts of the inner disk of LkCa15 (at least for the maximum particle sizes in the disk) while simultaneously the fragmentation probability for large particles in the inner parts is increased as well. If on the other hand the turbulence values were decreased, the picture changes slightly. Nonetheless, the model will still works since small $\alpha_{\rm turb}$ values benefit growth of very small particles to larger ones due to shifting of the sticking / bouncing transition to larger particle sizes. Since the collisional outcome model is somewhat pessimistic compared to other models (see \cite{Windmark2012a} for details), the transition of bouncing / mass transfer might be a priori shifted to larger sizes. The same is true for the disk HD135344B. Therefore, the recycling process can still be present for deviating turbulence values.

\section{CONCLUSION}

We calculated size and distance dependent drift velocities for dust particles in the inner dust ring of pre-transitional disks including photophoretic forces and radiation pressure. For the disks LkCa15 as well as HD135344B the resulting outer edge of the dust ring reachable by larger particles is in good agreement with the disk models (see fig. \ref{fig:drift3DLK}, \ref{fig:drift3DLKM2} and \ref{fig:drift3DHD}). 
Using the calculated velocities, we showed that a self sustained collisional recycling mechanism can be established in the inner dust ring of a pre-transitional disk.

\section*{ACKNOWLEDGMENTS}

This work is funded by the DFG under the grant number WU321/12-1, TE890/1-1 and DFG 1385. 
We thank the anonymous reviewer for a valuable review which was very helpful to improve this paper.

\bibliographystyle{aasjournal}
\bibliography{references}

\clearpage

\appendix

\section{Parameter sweep}

\begin{table*}
\caption{Parameter sweep for the disk LkCa15, $r_{\rm stop}$ marks the calculated maximum outward drift of a $10^{-4}$ m particle, $r_{\rm p,max}$ denotes the calculated maximum size where drift can still be positive.}             
\label{table:parametersweep}      
\centering                                      
\begin{tabular}{l l l l l | l l ||| l l l l l | l l}          
\hline\hline                        
$\delta_\dust$ 	&	$\delta_\gas$	&	$\chi_\dust$	& $k_{\rm th}$	& $\kappa_\nu$   	& \textbf{\bf$r_{\rm stop}$}	& \textbf{$r_{\rm p,max}$}		& $\delta_\dust$ 	&	$\delta_\gas$	&	$\chi_\dust$	& $k_{\rm th}$	& $\kappa_\nu$   	& \textbf{$r_{\rm stop}$}	& \textbf{$r_{\rm p,max}$}\\
&&&W/$\rm m\cdot K$&&AU&log($r_p / \rm m$)&&&&W/$\rm m\cdot K$&&AU&log($r_p / \rm m$)\\
\hline
0.00001	&	0.005	&	0.3	&	0.01	&	2.5	&	1.2	&	-2	&	0.00002	&	0.005	&	0.4	&	0.001	&	3	&	1.4	&	-2.4	\\
0.00001	&	0.005	&	0.3	&	0.01	&	3	&	1	&	-2.2	&	0.00002	&	0.005	&	0.4	&	0.001	&	3.5	&	1.2	&	-2.5	\\
0.00001	&	0.005	&	0.3	&	0.01	&	3.5	&	0.8	&	-2.4	&	0.00002	&	0.005	&	0.4	&	0.01	&	2.5	&	0.8	&	-2.5	\\
0.00001	&	0.005	&	0.4	&	0.01	&	2.5	&	1.5	&	-1.9	&	0.00002	&	0.005	&	0.5	&	0.001	&	3.5	&	1.8	&	-2.3	\\
0.00001	&	0.005	&	0.4	&	0.01	&	2.5	&	1.6	&	-1.9	&	0.00002	&	0.005	&	0.5	&	0.01	&	2.5	&	1	&	-2.2	\\
0.00001	&	0.005	&	0.4	&	0.01	&	3	&	1.2	&	-2	&	0.00002	&	0.005	&	0.5	&	0.01	&	3.5	&	0.7	&	-2.5	\\
0.00001	&	0.005	&	0.4	&	0.01	&	3.5	&	1.2	&	-2	&	0.00002	&	0.005	&	0.5	&	0.1	&	2.5	&	1	&	-2.2	\\
0.00001	&	0.005	&	0.5	&	0.01	&	2.5	&	1.7	&	-1.8	&	0.00002	&	0.01	&	0.3	&	0.001	&	2.5	&	1.6	&	-2.6	\\
0.00001	&	0.005	&	0.5	&	0.01	&	3	&	1.7	&	-2	&	0.00002	&	0.01	&	0.3	&	0.001	&	3	&	1	&	-2.8	\\
0.00001	&	0.005	&	0.5	&	0.01	&	3.5	&	1.5	&	-2	&	0.00002	&	0.01	&	0.4	&	0.01	&	2.5	&	1	&	-2.5	\\
0.00001	&	0.01	&	0.3	&	0.01	&	2.5	&	1.7	&	-1.9	&	0.00002	&	0.01	&	0.4	&	0.01	&	2.5	&	1	&	-2.4	\\
0.00001	&	0.01	&	0.3	&	0.01	&	3.5	&	1.2	&	-2.2	&	0.00002	&	0.01	&	0.5	&	0.01	&	2.5	&	1.5	&	-2.1	\\
0.00001	&	0.01	&	0.3	&	0.1	&	3	&	1.5	&	-2	&	0.00002	&	0.01	&	0.5	&	0.01	&	3	&	1.1	&	-2.3	\\
0.00001	&	0.01	&	0.4	&	0.01	&	3.5	&	1.8	&	-2	&	0.00002	&	0.01	&	0.5	&	0.01	&	3.5	&	1	&	-2.6	\\
0.00001	&	0.01	&	0.4	&	0.1	&	2.5	&	0.8	&	-2.1	&	0.00002	&	0.05	&	0.3	&	0.01	&	2.5	&	1.3	&	-2.8	\\
0.00001	&	0.01	&	0.4	&	0.1	&	3	&	0.8	&	-2.2	&	0.00002	&	0.05	&	0.4	&	0.01	&	3	&	1.5	&	-2.7	\\
0.00001	&	0.01	&	0.5	&	0.01	&	3.5	&	2	&	-1.9	&	0.00002	&	0.05	&	0.4	&	0.01	&	3.5	&	1.2	&	-2.9	\\
0.00001	&	0.01	&	0.5	&	0.1	&	2.5	&	1	&	-2	&	0.00002	&	0.05	&	0.5	&	0.01	&	3.5	&	1.8	&	-2.4	\\
0.00001	&	0.01	&	0.5	&	0.1	&	3	&	0.8	&	-2.1	&	0.00002	&	0.05	&	0.5	&	0.1	&	2.5	&	1	&	-2.8	\\
0.00001	&	0.05	&	0.3	&	0.1	&	2.5	&	1.2	&	-2.4	&	0.00002	&	0.05	&	0.5	&	0.1	&	3	&	0.8	&	-2.8	\\
0.00001	&	0.05	&	0.3	&	0.1	&	3	&	1.2	&	-2.6	&	0.00002	&	0.1	&	0.4	&	0.1	&	2.5	&	1	&	-2.9	\\
0.00001	&	0.05	&	0.3	&	0.1	&	3.5	&	0.8	&	-2.8	&	0.00003	&	0.005	&	0.4	&	0.001	&	2.5	&	1	&	-2.6	\\
0.00001	&	0.05	&	0.4	&	0.1	&	2.5	&	1.7	&	-2.2	&	0.00003	&	0.005	&	0.4	&	0.001	&	2.588	&	1	&	-2.6	\\
0.00001	&	0.05	&	0.4	&	0.1	&	3	&	1.5	&	-2.2	&	0.00003	&	0.005	&	0.5	&	0.001	&	2.5	&	1.5	&	-2.3	\\
0.00001	&	0.05	&	0.4	&	0.1	&	3.5	&	1.2	&	-2.5	&	0.00003	&	0.005	&	0.5	&	0.001	&	2.588	&	1.3	&	-2.4	\\
0.00001	&	0.05	&	0.5	&	0.1	&	2.5	&	1.8	&	-2	&	0.00003	&	0.005	&	0.5	&	0.001	&	3	&	1.05	&	-2.5	\\
0.00001	&	0.05	&	0.5	&	0.1	&	3	&	1.8	&	-2.1	&	0.00003	&	0.01	&	0.4	&	0.001	&	2.588	&	1.2	&	-2.7	\\
0.00001	&	0.05	&	0.5	&	0.1	&	3.5	&	1.7	&	-2.2	&	0.00003	&	0.01	&	0.5	&	0.001	&	2.588	&	2	&	-2.3	\\
0.00001	&	0.1	&	0.3	&	0.1	&	2.5	&	1.7	&	-2.4	&	0.00003	&	0.01	&	0.5	&	0.001	&	3	&	1.4	&	-2.5	\\
0.00001	&	0.1	&	0.3	&	0.1	&	3	&	1.6	&	-2.7	&	0.00003	&	0.01	&	0.5	&	0.001	&	3.5	&	1	&	-2.8	\\
0.00001	&	0.1	&	0.3	&	0.1	&	3.5	&	1.2	&	-2.8	&	0.00003	&	0.05	&	0.4	&	0.01	&	2.5	&	1	&	-3	\\
0.00001	&	0.1	&	0.4	&	0.1	&	3	&	2	&	-2.2	&	0.00003	&	0.05	&	0.4	&	0.01	&	2.588	&	1	&	-3	\\
0.00001	&	0.1	&	0.5	&	0.1	&	3.5	&	2	&	-2.2	&	0.00003	&	0.05	&	0.5	&	0.001	&	3.5	&	2	&	-3	\\
0.00002	&	0.005	&	0.3	&	0.001	&	2.5	&	1.3	&	-2.5	&	0.00003	&	0.05	&	0.5	&	0.01	&	2.588	&	1.3	&	-2.8	\\
0.00002	&	0.005	&	0.3	&	0.001	&	3	&	1	&	-2.7	&	0.00003	&	0.05	&	0.5	&	0.01	&	2.588	&	1.3	&	-2.7	\\
0.00002	&	0.005	&	0.4	&	0.001	&	2.5	&	2	&	-2.2	&	0.00003	&	0.05	&	0.5	&	0.01	&	3	&	1.05	&	-3	\\

\hline                                   
\hline
\end{tabular}
\end{table*}

\end{document}